\documentclass[reprint, aps, physrev, superscriptaddress, nofootinbib]{revtex4-2}

\usepackage{graphicx}
\usepackage{dcolumn}
\usepackage{bm}
\usepackage{amsmath}
\usepackage{amssymb}
\usepackage[utf8]{inputenc}
\usepackage[T1]{fontenc}
\usepackage{mleftright}
\usepackage{bbold}
\usepackage{braket}
\usepackage{mathtools}
\DeclarePairedDelimiter\abs{\lvert}{\rvert}%
\usepackage{hyperref}
\hypersetup{
  colorlinks   = true, 
  urlcolor     = blue, 
  linkcolor    = blue, 
  citecolor   = blue 
}

\usepackage{tikz}

\usepackage{mathptmx}

\def\e{\ensuremath{\mathrm{e}}}
\def\i{\ensuremath{\mathrm{i}}}
\def\d{\ensuremath{\mathrm{d}}}
\def\Tr{\ensuremath{\mathrm{Tr}}}

\newcommand{\kket}[1]{|#1\rangle\rangle}
\newcommand{\bbra}[1]{\langle\langle#1|}

\newcommand{\ant}[1]{\textcolor{black}{#1}}
\newcommand{\mar}[1]{\textcolor{black}{#1}}

\begin{document}

\title{Analytical solution of the open \ant{dispersive} Jaynes-Cummings model and \ant{perturbative analytical solution of the open} quantum Rabi model}
\date{\today}

\author{Antti Vaaranta}%
\email{antti.vaaranta@helsinki.fi}
\affiliation{QTF Centre of Excellence,  
Department of Physics, University of Helsinki, P.O. Box 43, FI-00014 Helsinki, Finland}
\affiliation{QTF Centre of Excellence, Department of Applied Physics,
		Aalto University, P.O. Box 15100, FI-00076 Aalto, Finland}
\author{Marco Cattaneo}%
\email{marco.cattaneo@helsinki.fi}
\affiliation{QTF Centre of Excellence,  
Department of Physics, University of Helsinki, P.O. Box 43, FI-00014 Helsinki, Finland}
\author{Paolo Muratore-Ginanneschi}
\email{paolo.muratore-ginanneschi@helsinki.fi}
\affiliation{Department of Mathematics and Statistics, University of Helsinki, P.O. Box 68, FI-00014 Helsinki, Finland}

\begin{abstract}
The Jaynes-Cummings and quantum Rabi models are fundamental to cavity and circuit quantum electrodynamics, as they describe the simplest form of light-matter interaction, where a single qubit is coupled to a single bosonic mode. A scenario that is commonly encountered in the experimental practice arises when the bosonic mode interacts with an external dissipative thermal bath, making the qubit-boson system open. 
In this work, we present new analytical solution of the Lindblad master equations for the open \ant{dispersive} Jaynes-Cummings \ant{model} and \ant{a perturbative analytical solution of the open} quantum Rabi model in the limit of weak qubit-boson coupling $g$, using the holomorphic formalism in Bargmann space. Specifically, we derive the most general solution of the local Lindblad master equation for the open dispersive Jaynes-Cummings model coupled to a thermal bath, with the only assumptions that the initial state of the qubit-boson system is separable. Additionally, we obtain a perturbative analytical solution for the open quantum Rabi model up to second order in $g$. Notably, our findings include a new formula for the qubit’s steady state at zeroth order, showing that the stationary populations depend on both qubit and boson frequencies in the quantum Rabi model, but not in the Jaynes-Cummings model, \mar{regardless of the value of $g$}. Our results are of general interest to the study of open quantum systems in the context of light-matter interaction.
\end{abstract}

\maketitle

\section{Introduction}

One of the simplest models for light-matter interaction is the Rabi model, developed originally by Isidor Rabi to study the effect of a classical magnetic field on an atom with nuclear spin \cite{Rabi1936, Rabi1937}. The fully quantized version of the original model, the quantum Rabi model, where the classical field was replaced by a quantized bosonic mode, was devised by Jaynes and Cummings \cite{Jaynes1963}. Despite its apparent simplicity, the analytical solution for the eigenenergies and eigenstates of the quantum Rabi model was found only as late as  2011 \cite{Braak2011}. Therefore, applying the rotating wave approximation to the quantum Rabi model, leading to the analytically simpler Jaynes-Cummings model, has been the go-to tool for obtaining a solvable, yet still accurate, description of light-matter interactions \cite{Larson2021, Braak2016,Larson2024,DeBernardis2024}.

In modern language, the Jaynes-Cummings and quantum Rabi models describe the coupling of a bosonic mode to a qubit (two-level quantum system)
, and are very relevant for the study and development of different platforms for quantum technologies. For example, they are broadly employed in cavity and circuit quantum electrodynamics (QED) \cite{Walther2006, Haroche2006, Blais2021}, where it is possible to carefully engineer systems in which a qubit interacts with a single mode of an electromagnetic field. In circuit QED the Jaynes-Cummings model is useful, for example, to study the readout of qubits, which is often realized by coupling a qubit to a coplanar waveguide that acts as a readout resonator with a single bosonic mode \cite{Blais2004, Wallraff2004}.

In the absence of any external interaction, both the quantum Rabi model and the Jaynes-Cummings model are closed quantum systems, where the unitary time evolution of the density operator is given by the Liouville-von Neumann equation. However, in many experimental realizations the bosonic mode is coupled to a thermal bath, making the qubit-boson system open \cite{Gambetta2006,Blais2021,Larson2021,Vaaranta2022, Lvov2024}. 
Assuming that the qubit-boson coupling is weak, the open system dynamics can be described through a \textit{local} master equation of the Lindblad form \cite{breuer2002theory, Trushechkin2016, Cattaneo2019}. \ant{Note that in this paper \mar{we say that the qubit-boson coupling $g$ is weak if it is much smaller than the qubit and boson frequencies}. This terminology does not necessarily conform with the one used in quantum optics, where for example $g=0.1$ would \mar{approach the  ``perturbative region of the ultra strong coupling regime'' \cite{Forn-Diaz2019}}, while here we still term it as ``weak''.}

Despite the relative simplicity of these models, few general analytical solutions of their local Lindblad master equations are known. In Ref.~\cite{PeixotodeFaria1999} the authors use superoperator methods in order to solve the local master equation for the dispersive Jaynes-Cummings model at zero temperature assuming that the initial state of the field is coherent. This solution is expanded to finite temperature in Ref.~\cite{Obada2008}. Coherent initial state and zero temperature limit are also considered in Ref.~\cite{Gambetta2006} using the P-representation method \cite{Walls2008}. In Ref.~~\cite{Clerk2007} the authors expand the study of the open dispersive Jaynes-Cummings model to finite temperatures using the Wigner function approach \cite{Wigner1932, Case2008, Schleich2011}, but assuming that the boson is initially in the thermal state. \mar{An analytical solution for the open quantum Rabi model with extremely large detuning and at zero temperature has been studied in Ref.~\cite{Hwang2018} using Keldysh path-integrals.}

In this paper, we present \mar{a new general analytical solution of the open dispersive Jaynes-Cummings model. \ant{Our only working hypothesis is that the initial state of the qubit-boson system is separable. Thus we expand the previous literature by dropping the assumption of an exact form of the initial state of the boson.} Moreover, we present a perturbative analytical solution of the quantum Rabi model, assuming that the initial state is separable and the boson is in a thermal state.} 

We solve the master equation using the 
so-called holomorphic formalism \cite{ZinnJustin2005, Hall2013, Vourdas1994, Vourdas2006, Hall2000}. Accordingly, we represent bosonic states by means of 
holomorphic functions 
of complex variables \cite{Bargmann1961, Bargmann1967}. 
We solve for the holomorphic representation of the propagation kernel, which determines the time evolution of an arbitrary initial state. 
This allows us to obtain a solution that is completely general and from which we can derive the evolution of thermal, coherent or other initial states of the boson and of an arbitrary initial state of the qubit.



In the case of the dispersive Jaynes-Cummings model we are able to solve the master equations exactly for the propagation kernels. \ant{The holomorphic formalism allows for an easy formulation of the general solution for arbitrary initial states as the effect of the initial state will be obtained via complex Gaussian integration.} Our general solution recovers the results of Refs.~\cite{Clerk2007,PeixotodeFaria1999} under the corresponding choice of the initial state of the boson. Additionally, we compare our analytical solutions to the numerical ones from Ref.~\cite{Vaaranta2022} and find an exact match between them. 



In the case of the quantum Rabi model, we apply perturbation theory around the qubit-boson coupling $g$ up to second order and away from resonance. There, we notice the appearance of resonances in the form of a secular (or time divergent) term, which is dealt with by applying multiscale perturbation theory \cite{Kevorkian1996,Verhulst2006,Strogatz2018}.
Moreover, we obtain a new formula for the zeroth-order steady state of the qubit. We observe that in the quantum Rabi model the stationary populations at zeroth order are not described by the standard Gibbs state of the qubit, but depend non-trivially on both the qubit and boson frequencies. The Gibbs state is instead recovered in the non-dispersive Jaynes-Cummings model. \ant{This steady state calculation is the only part of the paper where we consider the Jaynes-Cummings model without dispersive approximation.} We thus observe a non-perturbative difference in the steady state of these two models, which are usually considered equivalent in the regime of weak qubit-boson coupling \cite{Forn-Diaz2019,PhysRevA.86.033837,DeBernardis2024}. We once again compare the analytical results to numerics and find an excellent match for sufficiently small values of the qubit-boson coupling.

This article is organized as follows: in Section \ref{sec:physical-models} we introduce the Jaynes-Cummings and quantum Rabi models and their local master equations in the presence of a thermal bath. In Section \ref{sec:holom-form} we  briefly review  the holomorphic formalism and show the most useful identities needed to follow the logic of the calculations. In this section we also present the representation of the local master equations in the Bargmann space. The derivation of the expressions in this section and further details of the formalism are given in Appendix \ref{app:Holo}. Section \ref{sec:results} is dedicated to the analytical results of our work, while Section \ref{sec:conclusions} draws some concluding remarks.

\section{The physical models} \label{sec:physical-models}

\subsection{Relevant Hamiltonians}

The quantum Rabi model describes a physical system made of a single  qubit coupled to a single bosonic mode. Fixing $\hbar=1$, its Hamiltonian is defined as:
\begin{equation}
    \label{eqn:HamRabiModel}
    H_\text{Rabi} = \frac{\omega}{2} \sigma_z + \Omega a^\dagger a + g \sigma_x(a^\dagger +a ).
\end{equation}
In this work we are interested in the regime of weak qubit-boson interaction, and from now on we assume $g\ll \omega,\Omega$. 

Applying the rotating wave approximation to the quantum Rabi Hamiltonian, we obtain the Jaynes-Cummings model that is broadly employed in the description of light-matter interaction:
\begin{equation}
    \label{eqn:JCham}
    H_\text{JC}=\frac{\omega}{2} \sigma_z + \Omega a^\dagger a + g (\sigma_+ a+\sigma_- a^\dagger).
\end{equation}

While the expression for the eigenfunctions and eigenenergies of the Rabi model is quite involved \cite{Braak2011,Xie2017}, it can be easily shown that the infinite discrete energies of the Jaynes-Cummings model are \cite{Larson2021}:
\begin{equation}
    \label{eqn:energiesJC}
    E_\text{JC}(n,\pm)=\Omega\left(n\pm\frac{1}{2}\right)\pm\sqrt{\frac{\Delta^2}{4}+g^2(n+1)},
\end{equation}
where we have introduced the detuning $\Delta=\omega-\Omega$. The corresponding eigenvectors are
\begin{equation}
\label{eqn:eigenvectorsJC}
    \ket{\psi_{n,\pm}}=\sin\left(\frac{\theta_n}{2}\right)\ket{e,n}\pm \cos\left(\frac{\theta_n}{2}\right)\ket{g,n+1},
\end{equation}
where $\ket{g}$ and $\ket{e}$ are respectively the ground and excited states of the qubit, while
\begin{equation}
\label{eqn:arctanJC}
\theta_n=\arctan\left(\frac{2g\sqrt{n+1}}{\Delta}\right).
\end{equation}
If $\Delta =0$, then $\theta_n=\pi/2$ independently of $n$. 

\subsubsection{Dispersive regime of the Jaynes-Cummings model}
If the detuning $\Delta$ is much larger than the qubit-boson interaction $g$, then we are in the \textit{dispersive} regime of the Jaynes-Cummings model. The dispersive regime is widely employed in both cavity and circuit QED for the sake of qubit readout, qubit control, and many other tasks \cite{Haroche2006,Blais2021}. 

We introduce the perturbative parameter
\begin{equation}
\label{eqn:lambdaJC}
    \lambda=g/\Delta.
\end{equation}
Assuming $\abs{\lambda}\ll 1$,
we can transform the Jaynes-Cummings Hamiltonian Eq.~\eqref{eqn:JCham} into \cite{Zueco2009}:
\begin{equation}
    \label{eqn:dispersiveHam}
    H_\text{disp} = \frac{\omega'}{2}\sigma_z + \Omega a^\dagger a + g \lambda \sigma_z a^\dagger a,
\end{equation}
with the shifted qubit frequency $\omega'=\omega+g\lambda$. Eq.~\eqref{eqn:dispersiveHam} is accurate up to terms of the order $\mathcal{O}(\lambda^2)$. \ant{Note that the validity of the dispersive JCM relies also on the average photon number of the oscillator $\bar{n}$ to be small \cite{Blais2021}, which we take to hold for the examples considered in this text.}

While in the original Jaynes-Cummings Hamiltonian the qubit and the bosonic mode can exchange energy through the interaction term $\sigma_+ a + \sigma_{-} a^{\dagger}$ 
, in the dispersive regime this effect is negligible, i.e., the energy exchange  happens only at the order of $\mathcal{O}(\lambda^2)$. Then, the qubit-boson coupling in $H_\text{disp}$ is dispersive, meaning that the effective frequency of the bosonic mode is shifted depending on the state of the qubit, and there is no further interaction between the qubit and the boson in the first order of $\lambda$. This is why the dispersive regime is particularly important, for instance, for qubit readout in circuit QED \cite{Blais2021}.

\subsection{Open dynamics in the presence of a bath}
In this work, we are interested in the Jaynes-Cummings and Rabi models as open quantum systems. Specifically, we consider a scenario where the bosonic mode interacts with a local thermal bath that is not directly coupled to the qubit. This setup is of significant experimental relevance, as, for instance, it models light-matter interaction systems with lossy cavities \cite{Haroche2006}. In the context of circuit QED, such a configuration is particularly relevant for engineered dissipation schemes where a resistor is indirectly connected to the qubit via an intermediate resonator \cite{Tuorila2017,Cattaneo2021,Vaaranta2022}. This approach has been employed in several groundbreaking experiments in quantum thermodynamics \cite{Tan2017,Ronzani2018,Senior2020,Gubaydullin2022}.

The total Hamiltonian of the system-environment model can be written as:
\begin{equation}
\label{eqn:totHam}
    H_\text{tot}^{(\alpha)} = H_\alpha + \sum_k \omega_{\text{B},k} b_k^\dagger b_k + \sum_k \zeta_k \left(a b_k^\dagger+ a^\dagger b_k\right),
\end{equation}
with $\alpha=\text{Rabi},\text{JC},\text{disp}$, depending on the model we are studying. The set $\{b_k\}_k$ is an infinite collection of bosonic modes of the bath with frequency $\omega_{\text{B},k}$, which are coupled to the boson $a$ with coupling constant $\zeta_k$. 

Assuming a thermal bath with autocorrelation functions that decay fast in time, and a weak system-bath coupling constant, we can derive 
a Lindblad master equation for the reduced dynamics of the system. As is well known, solutions of such equation are always completely positive \cite{breuer2002theory}. The exact derivation of the master equation would require finding the dressed \textit{jump operators} of the qubit-boson system, which depend on the eigenstates of  $H_\alpha$ \cite{Carmichael1973,Cattaneo2019} (e.g., Eq.~\eqref{eqn:eigenvectorsJC} for the Jaynes-Cummings model). However, using the assumption of weak qubit-boson coupling $g$, we can derive an approximated \textit{local} master equation in which the jump operators are simply given by $a$ and $a^\dagger$ \cite{Trushechkin2016,Cattaneo2019}. 

The Lindblad master equation is
\begin{equation}
  \label{eq:masterEqLocal}
  \frac{\d}{\d t}\rho(t) = \mathcal{L}_\alpha[\rho(t)]\,,
\end{equation}
with
\begin{equation}
\label{eqn:masterEqSimple}
\begin{split}
    \mathcal{L}_\alpha[\rho]=& -i[H_\alpha,\rho]+\gamma(1+\bar{n})\left(a\rho a^\dagger-\frac{1}{2}\{a^\dagger a,\rho\}\right)\\
    &+\gamma\bar{n}\left(a^\dagger\rho a-\frac{1}{2}\{ aa^\dagger,\rho\}\right).
\end{split}
\end{equation}
Here, $\gamma$ is a decay rate that depends on the coupling constants $\zeta_k$ in Eq.~\eqref{eqn:totHam} \cite{breuer2002theory}, while $\bar{n}$ is the average number of photons according to the Bose-Einstein distribution:
\begin{equation}
\label{eqn:barNBoseEinstein}
    \bar{n}=\frac{1}{2}\left(\coth\left(\frac{\beta\Omega}{2}\right)-1\right),
\end{equation}
where $\beta$ is the inverse temperature of the thermal bath. We refer to $\mathcal{L}_\alpha$ as the \textit{Lindbladian} of the master equation.

Our goal is to find the general solution for $\rho(t)$ driven by $\mathcal{L}_\alpha$ in Eq.~\eqref{eqn:masterEqSimple}. We will focus on the quantum Rabi model ($\alpha=\text{Rabi}$) with system Hamiltonian given by Eq.~\eqref{eqn:HamRabiModel}, and on the dispersive Jaynes-Cummings model with Hamiltonian in Eq.~\eqref{eqn:dispersiveHam} ($\alpha=\text{disp}$). We will not generally treat $H_\text{JC}$ in Eq.~\eqref{eqn:JCham} as a separate case; however, we will consider it separately when determining the steady-state solutions. Indeed, the quantum Rabi model is more general than the Jaynes-Cummings model, and, at least for early times, we expect that their dynamics will be equivalent in the regime of small $g$, when the rotating wave approximation typically holds \cite{PhysRevA.86.033837,DeBernardis2024,Forn-Diaz2019}. Nonetheless, we will observe that the steady state solutions of these models can be different even at the zeroth perturbative order in $g$.

Note that, in the case of the Jaynes-Cummings model in the dispersive regime, the qubit operator $\sigma_z$ is a conserved quantity of the dynamics. Thus, the qubit energy is  not dissipated into the environment; instead, the thermal bath induces only dephasing in the qubit's state. In contrast, in the quantum Rabi model $[H_\text{Rabi},\sigma_z]\neq 0$, therefore the qubit dissipates energy into the thermal bath, even if it is not directly coupled to it.

\subsubsection{Steady state of the master equation}
\label{sec:steadyState}

In the case of the dispersive Jaynes-Cummings model, the local master equation Eq.~\eqref{eqn:masterEqSimple} has a 2-dimensional family of steady states due to the conserved quantity $\sigma_z$, which can be written as:
\begin{equation}
\label{eqn:ssSeparable}
    \rho_\text{ss}^{(\text{disp})}=(p\ket{g}\!\bra{g}+(1-p)\ket{e}\!\bra{e})\otimes\exp(-\beta \Omega a^\dagger a),
\end{equation}
for $p\in[0,1]$. The boson is driven towards its Gibbs state, while the qubit populations are conserved. In contrast, the qubit coherences vanish at late times. 

It is well-known that, in contrast to physical intuition, the steady state of the local master equation for the quantum Rabi (or non-dispersive Jaynes-Cummings) model is not the Gibbs state $\exp(-\beta H_\alpha)$ \cite{Cresser1992}. We may interpret this result as the presence of a physically relevant timescale, proportional to $\gamma^{-1}$, during which the local equation is valid and the state of the system is driven towards the non-Gibbs steady state of the local equation. However, in a much longer timescale (roughly captured by $\gamma^{-1}\omega/g$ \cite{Trushechkin2016}) we can expect the system to reach the Gibbs state predicted by statistical mechanics. This being said, the debate about the steady state issue is still open and different interpretations may be possible, depending also on the trade-off between $g,\gamma,$ and $\Delta$ \cite{Cattaneo2019,Trushechkin2021,Trushechkin2022}.

If the qubit and boson are decoupled ($g=0$), then the steady state of the boson is simply $\exp(-\beta \Omega a^\dagger a)$. In contrast, the qubit is not interacting with the thermal bath and the subspace of stationary states of the qubit-boson system is 2-dimensional: any state written as in Eq.~\eqref{eqn:ssSeparable} is a steady state. In this scenario, the qubit dynamics is not relaxing and depends on the initial conditions.  

Switching the qubit-boson interaction on ($g>0$) in the quantum Rabi and non-dispersive Jaynes-Cummings models, the qubit dynamics becomes dissipative and there is a sharp transition in the structure of the stationary subspace: there exists only a single stationary state of the qubit-boson system. If we expand this steady state as a function of $g$, we still recover a unique steady state at the zeroth order of $g$. In other words, only a single stationary state is ``selected'' in the family of Eq.~\eqref{eqn:ssSeparable} by the perturbative qubit-boson interaction, even at the lowest order of this perturbation. For the bosonic mode, this steady state must be the Gibbs state. In contrast, for the qubit this steady state cannot be deduced a priori, and we will observe that for the quantum Rabi model it does not correspond to the Gibbs state $\exp(-\beta\omega\,\sigma_z/2)$.

\section{Master equations in the Bargmann space}
\label{sec:holom-form}
The Hilbert space of the eigenstates of a quantum harmonic oscillator is usually represented as a Fock space, where the creation and annihilation operators of the harmonic oscillators are represented as infinite dimensional matrices. For this reason solving the Lindblad master equations in the Fock space is particularly difficult. Due to the infinite dimensionality, in numerical solutions one has to resort to truncation of the Hilbert space, as in Ref.~\cite{Vaaranta2022}, and obtaining analytical solutions directly in the Fock space is not feasible. To overcome the issue of dealing with infinite dimensional matrices, one can rely on the holomorphic formalism, passing from the Fock space to the Bargmann space \cite{Bargmann1961, Bargmann1967}. \ant{The holomorphic formalism is just a different way of representing the quantum states and operators \mar{of bosons}.}

In this section we give a brief overview of the holomorphic formalism that is used as a mathematical tool for obtaining analytical solutions of the open Jaynes-Cummings and quantum Rabi models discussed in the previous section. For completeness, here we just outline the basic ideas and identities of the method, which are more carefully derived in Appendix \ref{app:Holo} and discussed with detail in chapters 6 and 14 of \cite{ZinnJustin2005,Hall2013} respectively, as well as in Refs.~\cite{Vourdas1994, Vourdas2006, Hall2000}. Then, the presented identities are used for obtaining the Bargmann space representations for the local master equations describing the open quantum system dynamics.

\subsection{A brief review of the holomorphic formalism} 

\begin{figure}
  \centering
  \begin{tikzpicture}
    \node (fock)[rectangle, rounded corners=5pt, draw, align=left, label=Fock space] at (0,0) {
      $\begin{aligned}
        \ket{\phi} &= \sum_{n=0}^\infty \phi_n\ket{n} \\
        a &=
                  \begin{bmatrix}
                    0 & \sqrt{1} & 0 & \hdots \\
                    0 & 0 & \sqrt{2} & \\
                    0 & 0 & 0 & \\
                    \vdots &  &  & \ddots
                  \end{bmatrix} \\
        a\ket{n} &= \sqrt{n}\ket{n-1} \\
        \rho_\text{Th} &= \sum_{n=0}^\infty \left(\frac{\bar{n}}{1+\bar{n}}\right)^n\frac{\ket{n}\bra{n}}{1+\bar{n}}
      \end{aligned}
      $
    };
    \node (bargmann)[rectangle, rounded corners=5pt, draw, align=left, label=Bargmann space] at (5,0) {
      $\begin{aligned}
        f_\text{B}\left(\ket{\phi}; z\right) &= \sum_{n=0}^\infty\frac{\phi_nz^n}{\sqrt{n!}} \\
        F_\text{B}\left(a; z, z^*\right) &= z^*\e^{zz^*} \\
        f_\text{B}\left(a\ket{n}; z\right) &= nz^{n-1}\\
        F_\text{B}\left(\rho_\text{Th};z,z^*\right) &= \frac{\e^{\frac{\bar{n}}{1+\bar{n}}zz^*}}{1+\bar{n}}
      \end{aligned}
      $
    };
    \draw[->] (fock) -- node[above, align=center] {Coherent} (bargmann);
    \draw[->] (fock) -- node[below, align=center] {states} (bargmann);
  \end{tikzpicture}
  \caption{Examples of some objects in the Fock space and their corresponding holomorphic representations in the Bargmann space. The transformation between the two spaces is based on the coherent states $\ket{z}$. See Appendix \ref{app:Holo} for details.}
  \label{Fig:bridge-from-Fock-to-Bargmann}
\end{figure}
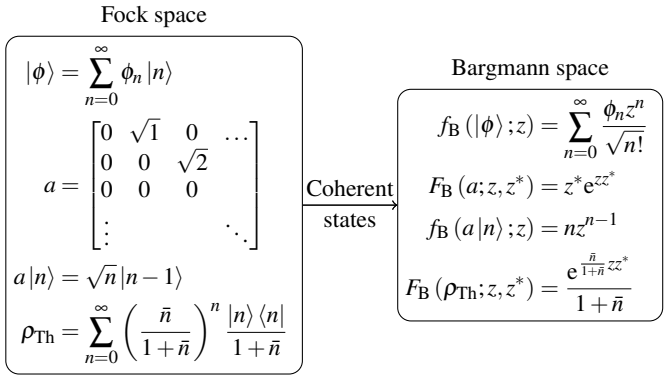

 The main ingredient for moving from the Fock space to the Bargmann space are the quantum optical coherent states $\ket{z}$, which are used to transform states in the  Fock space  and operators in the basis of the number of excitations  into holomorphic functions of complex variables in the Bargmann space.  This change of representation \ant{from Fock to Bargmann space} is depicted in Fig.~\ref{Fig:bridge-from-Fock-to-Bargmann}.

In the holomorphic formalism, the eigenstates of a quantum harmonic oscillator are represented as monomials of a single complex variable $z$ in the Bargmann space. For a general state $\ket{\phi}$ in the Fock space, the corresponding object in the Bargmann space is given by
\begin{equation}
  \label{eq:MainHoloState}
  f_\text{B}(\ket{\phi}; z)= \sum_{n=0}^\infty\phi_n\frac{z^n}{\sqrt{n!}}.
\end{equation}

Analogously, operators are represented as functions of $z$ and $z^*$ (which are independent variables in the complex plane) according to:
\begin{equation}
  \label{eq:MainHoloOp}
  F_\text{B}\left(O; z, z^*\right) = \sum_{n,m=0}^\infty O_{nm}\frac{z^nz^{*m}}{\sqrt{n!}\sqrt{m!}}\,,
\end{equation}
where $O_{nm}$ are the matrix elements of $O$ in the Fock basis. Moving operators back from the Bargmann space to the Fock space is done via the inverse map defined as
\begin{equation}
  \label{eq:operator-map-back-to-Fock}
  O = \frac{1}{\pi^2}\int\d^2z\d^2w \e^{-\frac{|z|^2}{2} - \frac{|w|^2}{2}}\ket{z^*}F_\text{B}\left(O; z, w^*\right)\bra{w}\,.
\end{equation}
Crucially, the action of an operator over a state in the holomorphic formalism is expressed through the integral
\begin{equation}
    \label{eq:MainOponState}
    f_\text{B}\left(O\ket{\psi}; z\right) = \frac{1}{\pi}\int \d^2w \e^{-|w|^2}F_\text{B}\left(O; z, w^*\right) f_\text{B}\left(\ket{\psi}; w\right)\,,
\end{equation}
where $F_\text{B}\left(O; z, z^*\right)$ acts as an integration kernel on the state function $f_\text{B}\left(\ket{\psi}; w\right)$ and the integration is over the whole space.

Linear transformations $S$ acting on operators $O$ and returning operators $S[O]$  are also represented as integration kernels. An example of this is the action of the Lindbladian $\mathcal{L}_\alpha$ 
on density matrices, giving the time evolved density matrix as
\begin{multline}
  \label{eq:time-evolved-rho-holo}
  F_\text{B}\left(\rho(t); z, z^*\right) = \frac{1}{\pi^2}\int\d^2w\d^2v\e^{-|w|^2-|v|^2}K(t;z,z^*|w,v)\\
  \times F_\text{B}(\rho(0);w^*,v^*)\,,
\end{multline}
where $K(t;z,z^*|w,v)$ is the propagation kernel arising from   $\mathcal{L}_\alpha$. \ant{Its exact form depends on the master equation at hand, and a concrete example will be given in the next section.} \ant{The propagation kernel $K(t;z,z^*|w,v)$} evolves the initial state $F_\text{B}(\rho(0);w^*,v^*)$ from time $0$ to $t$ via integration over the auxiliary complex variables $w,v$.


\ant{
\subsubsection{Transformation of the ladder operators}
}
\ant{
The holomorphic formalism is most useful when dealing with harmonic oscillators, which naturally give rise to creation and annihilation operators. By the direct use of the Fock-to-Bargmann transformation for operators (\ref{eq:MainHoloOp}), it can be shown that the ladder operators transform as follows:
\begin{subequations}
  \begin{align}
    a &\to F_\text{B}\left(a; z, z^*\right) = z^*\e^{zz^*} \,, \label{eq:annihilation-op-holomorphic} \\
    a^\dagger &\to F_\text{B}\left(a^\dagger; z, z^*\right) = z\e^{zz^*} \,. \label{eq:creation-op-holomorphic}
  \end{align}
\end{subequations}
These functions will be used as integration kernels and the action of these can be computed by taking some holomorphic test function $f_\text{B}(\ket{\psi}; z)$, multiplying it by the corresponding integration kernel and integrating over the whole complex space as defined in Eq.~(\ref{eq:MainOponState}).
Performing this operation for both (\ref{eq:annihilation-op-holomorphic}) and (\ref{eq:creation-op-holomorphic}) yields
\begin{subequations}
  \begin{align}
    a\ket{\psi} &\to f_\text{B}(a\ket{\psi}; z) = \frac{\d}{\d z}f_\text{B}(\ket{\psi}; z) \,, \label{eq:action-of-annihilation-op-holomorphic} \\
    a^\dagger\ket{\psi} &\to f_\text{B}(a^\dagger\ket{\psi}; z) = zf_\text{B}(\ket{\psi}; z) \,. \label{eq:action-of-creation-op-holomorphic}
  \end{align}
\end{subequations}
We can notice that in the holomorphic formalism the action of the creation and annihilation operators can be replaced by multiplication and differentiation with the complex variable respectively: $a^\dagger \leftrightarrow z\,,a\leftrightarrow\d/\d z$.
}

\ant{
 When dealing with master equations, we often come up with expressions where the density operator $\rho$ is acted upon by the ladder operators $a$ and $a^\dagger$ from the left, right or from both directions. Applying the definition on how operators are mapped to the Bargmann space in Eq.~(\ref{eq:MainHoloOp}), we can easily derive the following transformation rules:}

\ant{
\begin{subequations}
 \begin{minipage}{0.45\linewidth}
   \begin{align}
     a_\text{left} &\leftrightarrow \frac{\partial}{\partial z}\,, \label{eq:a_L-map} \\
     a_\text{left}^\dagger &\leftrightarrow z\,, \label{eq:a_L^dag-map}
   \end{align}
 \end{minipage}\hfill
 \begin{minipage}{0.45\linewidth}
   \begin{align}
     a_\text{right} &\leftrightarrow z^*\,, \label{eq:a_R-map}\\
     a_\text{right}^\dagger &\leftrightarrow \frac{\partial}{\partial z^*}\,. \label{eq:a_R^dag-map}
   \end{align}
\end{minipage}
\end{subequations}
\vspace{7pt}
}

\ant{
  \noindent These rules encode the difference between left and right operations to the complex variable being conjugated or not.
  }
                                            

\subsection{The Bargmann space representation of the local master equations} \label{sec:local-mast-equat}

In order to solve the local master equations governing the open dynamics of the dispersive Jaynes-Cummings and quantum Rabi Hamiltonians (Eqs.~(\ref{eqn:dispersiveHam}) and (\ref{eqn:HamRabiModel}) respectively), we now introduce the Bargmann space representation of Eq.~(\ref{eq:masterEqLocal}). This allows us to transform the effectively infinite system of linear equations into four (possibly coupled) complex partial differential equations (PDEs).

First, we apply Eq.~(\ref{eq:time-evolved-rho-holo}) to the time evolved density matrix \ant{together with the identities (\ref{eq:a_L-map}) to (\ref{eq:a_R^dag-map})}. Taking into account the structure of the two-dimensional Hilbert space of the qubit, we write the Bargmann space density matrix as
\begin{multline}
  \label{eq:rho-in-holomorphic_main}
  F_\text{B}\left(\rho(t); z, z^*\right) = \frac{1}{\pi^2}\int\d^2w\d^2v\e^{-|w|^2-|v|^2}
    \begin{bmatrix}
      A(t) & B(t) \\
      C(t) & D(t)
    \end{bmatrix}\\
  \times F_\text{B}\left(\rho_\text{f}(0); w^*, v^*\right)\,,
\end{multline}
where $F_\text{B}\left(\rho_\text{f}(0); w^*, v^*\right)$ is the initial state of the boson. Here we have assumed that the initial state of the system in factorized, such that $\rho(0) = \rho_\text{Q}(0)\otimes\rho_\text{f}(0)$. Generalizations to separable states are trivial, as the state evolution is described by a linear map. We write the initial state of the qubit as 
\begin{equation}
\label{eqn:matrixInStateQ}
    \rho_\text{Q}(0)=\begin{bmatrix}
        q_{11} & q_{12}\\ 
        q_{21} & q_{22}
    \end{bmatrix},
\end{equation}
where $q_{11}$ ($q_{22}$) is the population of the excited (ground) state, $q_{11} + q_{22} = 1$, and $q_{21} = q_{12}^*$.

  The functions $A, B, C$ and $D$ can be obtained through matrix multiplication between the propagation kernel $K(t;z,z^*|w,v)$, which we choose to write as a $2\times 2$ matrix because of the structure of the qubit Hilbert space, and the initial density matrix of the qubit:
\begin{subequations}
  \begin{align}
    A(t) &\equiv q_{11}K_{11}(t; z,z^*|w,v) + q_{21}K_{12}(t; z,z^*|w,v)\,, \label{eq:A(t)}\\
    B(t) &\equiv q_{12}K_{11}(t; z,z^*|w,v) + q_{22}K_{12}(t; z,z^*|w,v)\,, \label{eq:B(t)}\\
    C(t) &\equiv q_{11}K_{21}(t; z,z^*|w,v) + q_{21}K_{22}(t; z,z^*|w,v)\,, \label{eq:C(t)}\\
    D(t) &\equiv q_{12}K_{21}(t; z,z^*|w,v) + q_{22}K_{22}(t; z,z^*|w,v)\,, \label{eq:D(t)}
  \end{align}
\end{subequations}
where $K_{ij}$ are the matrix element of $K(t;z,z^*|w,v)$. Note that, in order to make the notation shorter, we keep only the time dependency in the above definitions. However, these functions depend also on some complex variables: $A(t) \equiv A(t; z,z^*|w,v)$ and analogously for the other functions.

The initial condition of the propagation kernel
\begin{equation}
  \label{eq:propagation-kernel-ics}
  K(0; z, z^*|w, v) =
  \begin{bmatrix}
    1 & 0 \\
    0 & 1
  \end{bmatrix}
  \e^{zw + z^*v}
\end{equation}
is equivalent to the identity operators in the qubit space (the identity matrix) and in the Bargmann space (the exponential kernel). This initial condition then determines the initial conditions for the propagator kernel functions $A(t)$ to $D(t)$, which will be needed for solving the arising PDEs.


Applying Eq.~(\ref{eq:rho-in-holomorphic_main}) to the case of the dispersive Jaynes-Cummings model Hamiltonian in Eq.~(\ref{eqn:dispersiveHam}), the corresponding local master equation (\ref{eq:masterEqLocal}) transforms into a set of four uncoupled PDEs for the propagator kernel functions:
\begin{subequations}
  \begin{align}
      \frac{\d}{\d t}A(t) &= \mathcal{D}^{(+,+)}(z,z^*)A(t)\,, \label{eq:de-for-A(t)}  \\
      \frac{\d}{\d t}B(t) &= \left[\mathcal{D}^{(+,-)}(z,z^*) - \i\omega^\prime\right]B(t)\,, \label{eq:de-for-B(t)} \\
      \frac{\d}{\d t}C(t) &= \left[\mathcal{D}^{(-,+)}(z,z^*) + \i\omega^\prime\right]C(t)\,, \label{eq:de-for-C(t)}  \\     
      \frac{\d}{\d t}D(t) &= \mathcal{D}^{(-,-)}(z,z^*)D(t)\,, \label{eq:de-for-D(t)}
  \end{align}
\end{subequations}
state of the qubit+boson system; ii) a perturbative analytical solution of the QRM up to the second order inwhere $\mathcal{D}^{(i,j)}(z,z^*)$ is a differential operator defined as
\begin{multline}
  \label{eq:diff-op-DFirst}
  \mathcal{D}^{(i,j)}(z,z^*) = -\phi_i z\frac{\partial}{\partial z} - \phi_j^* z^*\frac{\partial}{\partial z^*} + \gamma(\bar{n}+1)\frac{\partial}{\partial z}\frac{\partial}{\partial z^*} \\
  + \gamma\bar{n}zz^* - \gamma\bar{n}\,,
\end{multline}
and $i,j\in\{+,-\}$ with $\phi_\pm = \gamma(\bar{n} + 1/2) + \i(\Omega \pm g\lambda)$.

In the case of the quantum Rabi model Hamiltonian in Eq.~(\ref{eqn:HamRabiModel}), the PDEs corresponding to the local master equation in the Bargmann space are instead coupled to each other:
\begin{subequations}
  \begin{align}
    \begin{split}
      \frac{\d}{\d t}A(t) &= \mathcal{D}(z,z^*)A(t) \\
      &\quad + \i g\bigg[\left(z^*+\frac{\partial}{\partial z^*}\right)B(t) - \left(z+\frac{\partial}{\partial z}\right)C(t)\bigg]\,, \label{eq:Rabi-DE-A}
    \end{split}
    \\
    \begin{split}
      \frac{\d}{\d t}B(t) &= \left[\mathcal{D}(z,z^*) - \i\omega\right]B(t) \\
      &\quad + \i g\bigg[\left(z^*+\frac{\partial}{\partial z^*}\right)A(t) - \left(z+\frac{\partial}{\partial z}\right)D(t)\bigg]\,, \label{eq:Rabi-DE-B}
    \end{split}
    \\
    \begin{split}
      \frac{\d}{\d t}C(t) &= \left[\mathcal{D}(z,z^*) + \i\omega\right]C(t) \\
      &\quad + \i g\bigg[\left(z^*+\frac{\partial}{\partial z^*}\right)D(t) - \left(z+\frac{\partial}{\partial z}\right)A(t)\bigg]\,, \label{eq:Rabi-DE-C}
    \end{split}
    \\
    \begin{split}
      \frac{\d}{\d t}D(t) &= \mathcal{D}(z,z^*)D(t) \\
      &\quad + \i g\bigg[\left(z^*+\frac{\partial}{\partial z^*}\right)C(t) - \left(z+\frac{\partial}{\partial z}\right)B(t)\bigg]\,. \label{eq:Rabi-DE-D}
    \end{split}
  \end{align}
\end{subequations}
Here $\mathcal{D}(z,z^*)$ is a differential operator that has the form
\begin{multline}
  \label{eq:diff-op-DSecond}
  \mathcal{D}(z,z^*) = -\phi z\frac{\partial}{\partial z} - \phi^* z^*\frac{\partial}{\partial z^*} + \gamma(\bar{n}+1)\frac{\partial}{\partial z}\frac{\partial}{\partial z^*} \\
  + \gamma\bar{n}zz^* - \gamma\bar{n}\,,
\end{multline}
where $\phi = \gamma(\bar{n} + 1/2) + \i\Omega$. 

\section{Results} \label{sec:results}
 We are now ready to present the solution to the PDEs in Bargmann space that govern the open system dynamics. For the dispersive Jaynes-Cummings model the uncoupled differential equations (\ref{eq:de-for-A(t)}) to (\ref{eq:de-for-D(t)}) can be solved exactly via a Gaussian ansatz. However, for the quantum Rabi model the situation is not so fortunate as the PDEs are coupled. In this case we resort to multiscale perturbation theory up to second order in the small qubit-boson coupling parameter $g$.

It should be noted that solving the PDEs (\ref{eq:de-for-A(t)}) to (\ref{eq:de-for-D(t)}) and (\ref{eq:Rabi-DE-A}) to (\ref{eq:Rabi-DE-D}) gives solutions to the propagation kernels $A-D$, not to the quantum system dynamics directly. This has the benefit that we are not restricting the form of the initial state of the system in any way during the process. Therefore, we can find the solution for the time evolution of any initially separable qubit-boson state. This computation requires finding the holomorphic representation of the boson's initial state, multiplying it with the propagation kernel and integrating over the extra complex variables as shown in Eq.~(\ref{eq:time-evolved-rho-holo}). After obtaining the dynamics of the system in the Bargmann space, one can apply the inverse mapping (\ref{eq:operator-map-back-to-Fock}) to move back to the more familiar Fock space.

In this section we first consider the dispersive Jaynes-Cummings model before moving to the more general quantum Rabi model. In each subsection we first discuss the solution for the propagators in general, and then show as an example how the obtained solution can be used to compute the time evolution for some specific initial state of the boson.

\subsection{Open dispersive Jaynes-Cummings model}
\label{sec:openDis}

The solution to each of the differential equations (\ref{eq:de-for-A(t)}) to (\ref{eq:de-for-D(t)}) can be obtained by applying a Gaussian ansatz for the propagator kernels. For example, for $A(t)$ we would write
\begin{equation}
  \label{eq:Gauss-ansatz}
  A(t) \equiv A(t;z,z^*|w,v) = Z^{(A)}(t)\e^{\bm{x}^T\bm{H}^{(A)}(t)\bm{x}}\,,
\end{equation}
where $\bm{x} = [z,z^*,w,v]^T$ is a vector of the complex variables and $\bm{H}^{(A)}(t)$ is some matrix with each of its elements $H^{(A)}_{ij}(t)$ being a function of time only. The ansatz has an initial condition $A(0;z,z^*|w,v) = q_{11}\e^{zw+z^*v}$, such that at time $t=0$ it must equal the identity kernel times $q_{11}$, which is the initial population of the excited state of the qubit according to Eq.~\eqref{eqn:matrixInStateQ}.

Applying the ansatz (\ref{eq:Gauss-ansatz}) to equation (\ref{eq:de-for-A(t)}) we find a bunch of differential equations for the coefficient $Z^{(A)}(t)$ and the  matrix elements $H^{(A)}_{ij}(t)$. These differential equations can be solved using standard methods and finally we obtain the following results in the case of $A(t;z,z^*|w,v)$:
\begin{equation}
  Z^{(A)}(t) = \frac{q_{11}}{1 + \bar{n}(1 - \e^{-\gamma t})}\, ,
\end{equation}
and the matrix $\bm{H}^{(A)}(t)$ can be written as
\begin{widetext}
  \begin{equation}
    \label{eq:H(t)-matrix}
    \bm{H}^{(A)}(t) = \frac{1}{2 + 2\bar{n}(1 - \e^{-\gamma t})}
    \begin{bmatrix}
      0 & \bar{n}(1 - \e^{-\gamma t}) & \e^{-\gamma t/2}\e^{-\i(\Omega + g\lambda)t} & 0 \\
      \bar{n}(1 - \e^{-\gamma t}) & 0 & 0 & \e^{-\gamma t/2}\e^{\i(\Omega + g\lambda)t} \\
      \e^{-\gamma t/2}\e^{-\i(\Omega + g\lambda)t} & 0 & 0 & (\bar{n} + 1)(1 - \e^{-\gamma t}) \\
      0 & \e^{-\gamma t/2}\e^{\i(\Omega + g\lambda)t} & (\bar{n} + 1)(1 - \e^{-\gamma t}) & 0
    \end{bmatrix}\,.
  \end{equation}
\end{widetext}
It should be noted that this solution satisifes the initial conditions because $Z^{(A)}(0) = q_{11}$ and $H^{(A)}_{ij} = 0$ for all $i,j$ except for $H^{(A)}_{13}(0) = H^{(A)}_{24}(0) = H^{(A)}_{31}(0) = H^{(A)}_{42}(0) = 1/2$, which gives then $\e^{\bm{x}^T\bm{H}^{(A)}(0)\bm{x}} = \e^{zw+z^*v}$.

For $B(t;z,z^*|w,v)$ the situation is similar but the differential equation (\ref{eq:de-for-B(t)}) to be solved changes slightly. However, this does not affect the form of the ansatz that solves the PDE, so we use the same form as in Eq.~(\ref{eq:Gauss-ansatz}) but with corrected initial data $B(0;z,z^*|w,v) = q_{12}\e^{zw+z^*v}$. Again we get differential equations for $H^{(B)}_{ij}(t)$ and $Z^{(B)}(t)$, which can be solved to yield
\begin{equation}
  \label{eq:Z^B(t)}
  Z^{(B)}(t) = \frac{q_{12}\e^{(\theta_1 - \theta_2)t/2}}{\e^{-\theta_2t} + C_1\left(1 - \e^{-\theta_2t}\right)}\,,
\end{equation}
where $C_1$ is given by
\begin{equation}
  \label{eq:C-coeff-B(t)}
  C_1 = \frac{\gamma\bar{n}+\i\omega^\prime}{\theta_2} + \frac{\theta_1}{2\theta_2} + \frac{1}{2}
\end{equation}
and $\theta_1, \theta_2$ are
\begin{subequations}
    \begin{align}
      \label{eq:theta1}
      \theta_1 &= \gamma - 2\i(\omega^\prime - g\lambda)\,, \\
      \label{eq:theta2}
      \theta_2 &= \sqrt{\gamma^2 - 4g^2\lambda^2 + 2\i\gamma(1 + 2\bar{n})2g\lambda}\,.
    \end{align}
\end{subequations}
The matrix $\bm{H}^{(B)}(t)$ is written as
\begin{widetext}
  \begin{equation}
    \label{eq:H(t)-matrix-B}
    \bm{H}^{(B)}(t) = \frac{1}{2\e^{-\theta_2 t} + 2C_1(1 - \e^{-\theta_2t})}
    \begin{bmatrix}
      0 & \frac{\theta_2C_1(C_1-1)(1 - \e^{-\theta_2t})}{\gamma(\bar{n}+1)} & \e^{-\i\Omega t}\e^{-\theta_2t/2} & 0 \\
      \frac{\theta_2C_1(C_1-1)(1 - \e^{-\theta_2t})}{\gamma(\bar{n}+1)} & 0 & 0 & \e^{\i\Omega t}\e^{-\theta_2t/2} \\
      \e^{-\i\Omega t}\e^{-\theta_2t/2} & 0 & 0 & \frac{\gamma(\bar{n} + 1)(1 - \e^{-\theta_2t})}{\theta_2} \\
      0 & \e^{\i\Omega t}\e^{-\theta_2t/2} & \frac{\gamma(\bar{n} + 1)(1 - \e^{-\theta_2t})}{\theta_2} & 0
    \end{bmatrix}\,.
  \end{equation}
\end{widetext}

It can be shown that $\text{Re}[\theta_2]\geq\gamma$, which makes sure that Eq.~(\ref{eq:Z^B(t)}) does not diverge in time. The proof of this relies in noticing that $\text{Re}[\theta_2]$ is a monotonic and increasing function, with a derivative that is always positive. Thus the expression obtains its minimum at the boundaries of its domain. This minimum can then be shown to equal $\gamma$.

The form of the remaining elements $C(t;z,z^*|w,v)$ and $D(t;z,z^*|w,v)$ can be deduced from the solutions for $A(t;z,z^*|w,v)$ and $B(t;z,z^*|w,v)$ as there are only minor changes in the signs of the coefficients in the underlying differential equations (\ref{eq:de-for-C(t)}) and (\ref{eq:de-for-D(t)}). For $D(t;z,z^*|w,v)$ we must change $\phi_+ \to \phi_-$, which only amounts to changing the sign of the $+g\lambda$ terms in Eq.~(\ref{eq:H(t)-matrix}). We can also notice that the PDE for $C(t;z,z^*|w,v)$ (Eq.~(\ref{eq:de-for-C(t)})) is the complex conjugate of that for $B(t;z,z^*|w,v)$, so the solution is also just the conjugate of the one from Eqs.~(\ref{eq:Z^B(t)}) to (\ref{eq:H(t)-matrix-B}).

\subsubsection{Dynamics with thermal initial condition} \label{sec:dynam-with-therm}
We have now solved for the propagation kernels $A(t;z,z^*|w,v)$ to $D(t;z,z^*|w,v)$. Using Eq.~(\ref{eq:rho-in-holomorphic_main}) we are able to compute the exact dynamics, once we specify the initial state $F_\text{B}(\rho_\text{f}(0); z,z^*)$ of the boson. Let us first take the boson to be in the thermal state, which in the holomorphic formalism is written as
\begin{equation}
  \label{eq:thermal-state-holo}
  F_\text{B}(\rho_\text{Th};z,z^*) = \frac{1}{1+\bar{n}}\e^{\frac{\bar{n}}{\bar{n}+1}zz^*}\,.
\end{equation}
After performing the complex Gaussian integration for each of the matrix elements,  we obtain the solution in the Bargmann space as
\begin{equation}
  \label{eq:holomorphic-solution-rho(t)}
  F_\text{B}^{(i)}\left(\rho(t); z, z^*\right) = \frac{Z^{(i)}(t)}{1 + \bar{n} - 2\bar{n}H^{(i)}_{34}(t)}\e^{\left(\frac{4\bar{n}H^{(i)}_{13}(t)H^{(i)}_{24}(t)}{1 + \bar{n} - 2\bar{n}H^{(i)}_{34}(t)} + 2H^{(i)}_{12}(t)\right)zz^*}\,,
\end{equation}
where $i$ refers to the $A,B,C$ or $D$ quadrant of the matrix and the functions $H_{ij}(t)$ are given in equations (\ref{eq:H(t)-matrix}) and (\ref{eq:H(t)-matrix-B}). Note that the denominators here and in the preceding sections of the paper are always well defined, which can easily be verified by using the expressions for the functions $H_{ij}(t)$ from Eqs.~(\ref{eq:H(t)-matrix}) and (\ref{eq:H(t)-matrix-B}).

The last step in the process is to transform the obtained Bargmann space result back to the Fock space for an easier interpretation. This is done by applying the backwards transformation from Eq.~(\ref{eq:operator-map-back-to-Fock}), which gives once again a complex Gaussian integral. Performing the integral gives us the matrix elements of the density matrix as
\begin{multline}
  \label{eq:rho-solution-Fock}
  \rho^{(i)}(t) = \frac{Z^{(i)}(t)}{1 + \bar{n} - 2\bar{n}H^{(i)}_{34}(t)}\\
  \sum_{n=0}^\infty\left(\frac{4\bar{n}H^{(i)}_{13}(t)H^{(i)}_{24}(t)}{1 + \bar{n} - 2\bar{n}H^{(i)}_{34}(t)} + 2H^{(i)}_{12}(t)\right)^n\ket{n}\bra{n}\,.
\end{multline}
This solution contains both the qubit and boson density matrices. Taking the trace over the bosonic degrees of freedom and performing the resulting geometric sum (the summation is well defined as the summand is always less than unity) gives us the elements of the qubit density matrix as
\begin{equation}
  \label{eq:qubit-dm}
  \rho_\text{Q}^{(i)}(t) = \frac{Z^{(i)}(t)}{\splitfrac{\left(1 + \bar{n} - 2\bar{n}H^{(i)}_{34}(t)\right)\left(1 - 2H^{(i)}_{12}(t)\right)}{ - 4\bar{n}H^{(i)}_{13}(t)H^{(i)}_{24}(t)}}\,.
\end{equation}
The above expressions are the qubit density matrix elements in the usual excited--ground state basis, where $i=A$ refers to the population of the excited state, $i=D$ to  that of the ground state, and $i=B,C$ to the qubit coherences. Plugging in the obtained results for the different functions $H^{(i)}(t)$ and $Z^{(i)}(t)$ gives $\rho^{(A)}(t) = q_{11}$ and $\rho^{(D)}(t) = q_{22}$ as it should, since the dispersive Jaynes-Cummings model experiences no dissipation.

This analytical result allows us to compute for example the time evolution of the expectation values of the Pauli operators as $\langle\sigma_{x,y,z}\rangle = \Tr[\sigma_{x,y,z}\rho_\text{Q}(t)]$. Then, we can compute the coherence measure
\begin{equation}
  \label{eq:coh-measure}
  C(\rho_\text{Q}(t)) = \sqrt{\langle\sigma_x\rangle^2 + \langle\sigma_y\rangle^2}\,,
\end{equation}
which quantifies the amount of coherence of the qubit state at certain time \cite{Baumgratz2014}.

\begin{figure*}
  \centering
  \includegraphics[width=\textwidth]{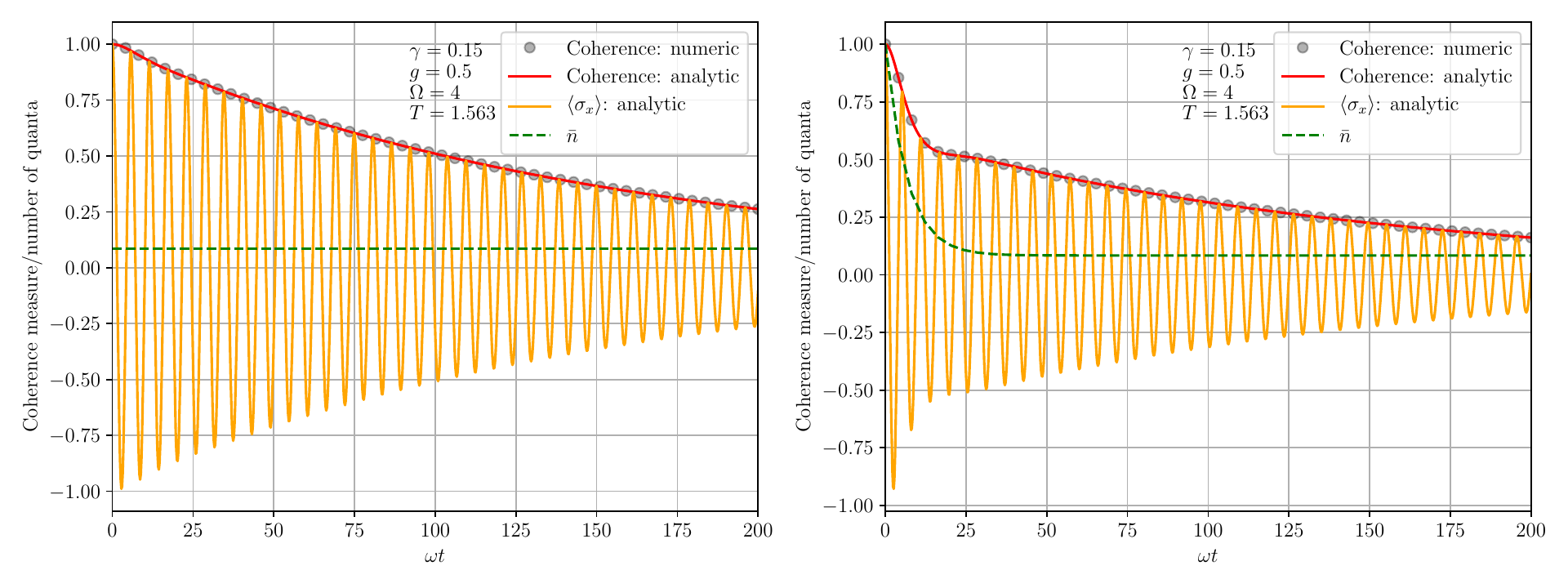}
  \caption{Evolution of the coherence measure for the qubit and of the average number of quanta in the bosonic mode as a function of normalized time in the open dispersive Jaynes-Cummings model. The parameters are chosen as in Ref.~\cite{Vaaranta2022}: $\omega=1,\ \gamma = 0.15,\ g=0.5,\ \Omega = 4$ and $T=1.563$. The qubit is initialized in the superposition state $\ket{+} = 1/\sqrt{2}(\ket{g} + \ket{e})$. Left: the boson is initialized in the thermal state with temperature $T$ (compare with Fig.~8 of \cite{Vaaranta2022}). Right: the boson is initialized in a coherent state (compare with Fig.~9 of \cite{Vaaranta2022}). The grey dots capture the coherence measure in Eq.~(\ref{eq:coh-measure}) computed numerically, while the red solid line shows its analytical prediction obtained through Eq.~\eqref{eq:qubit-dm} (left) and Eq.~\eqref{eq:qubit-dm-coherent-ics} (right). The coherence measure is the envelope of the time evolution of $\langle\sigma_x\rangle$, which is computed analytically (oscillating orange solid line). The green dashed line is the expected number of excitations in the boson.}
  \label{fig:coherence-evol}
\end{figure*}

Next we compare the analytical results to numerics obtained by solving the local master equation using QuTiP \cite{Johansson2012,Johansson2013}. The parameters are set to be the same as in Ref.~\cite{Vaaranta2022}, namely $\omega = 1, \Omega = 4, \gamma = 0.15, g=0.5$ and $T=1.563$. The results of the comparison are shown in the left panel of Fig.~\ref{fig:coherence-evol}, where we see that the correspondence between the analytics and numerics is exact. In the numerics the boson Hilbert space was truncated to 7 levels, with the highest level having a population of $10^{-7}$. Therefore, our analytical results also provide a benchmark for the truncation level that is needed to obtain accurate numerical simulations.

As the dynamics of the qubit coherences are determined by the off-diagonal elements of the density matrix, we can deduce from Eq.~(\ref{eq:qubit-dm}) that the driving term for the coherence decay is $Z^{(B)}(t)$ (and its complex conjugate $Z^{(C)}(t)$). Then, from Eq.~(\ref{eq:Z^B(t)}), we note that the rate of coherence decay is determined by the argument of the exponential, namely
\begin{equation}
  \label{eq:rate-of-coherence-decay}
  \Gamma_2 = \frac{1}{2}\text{Re}\left[\gamma - \sqrt{\gamma^2 - 4g^2\lambda^2 + 2\i\gamma(1 + 2\bar{n})2g\lambda}\right]\,.
\end{equation}
We can therefore relate this expression to the slowest decaying eigenmode of the Liouvillian block that contributes to decoherence, as was done in Ref.~\cite{Vaaranta2022}. Moreover, this expression is equivalent to Eq.~(44) in Ref.~\cite{Clerk2007}, which was obtained through an approach based on the Wigner function representation.

\subsubsection{Dynamics with coherent initial condition}
Let us now assume that the boson is initially in a coherent state. In the Bargmann space the coherent state can be written as
\begin{equation}
  \label{eq:coherent-dm-holomorphic}
  F_\text{B}\left(\rho_\alpha; z, z^*\right) = \e^{-|\alpha|^2 + z\alpha + z^*\alpha^*}\,.
\end{equation}
Next, using Eq.~(\ref{eq:rho-in-holomorphic_main}), with the propagation kernels given as the Gaussians like in Eq.~(\ref{eq:Gauss-ansatz}), we find the time evolved density operator in Bargmann space as
\begin{multline}
  \label{eq:rho-holomorphic-coherent-ics}
  F_\text{B}^{(i)}\left(\rho(t); z, z^*\right) = Z^{(i)}(t)\exp\Big(2H^{(i)}_{12}(t)zz^* + 2H^{(i)}_{13}(t)\alpha z \\
  + 2H^{(i)}_{24}(t)\alpha^*z^* - |\alpha|^2(1 - 2H^{(i)}_{34}(t))\Big)\,.
\end{multline}
Then we can convert this expression back to the Fock space using Eq.~(\ref{eq:operator-map-back-to-Fock}), compute the integral, and take the trace over the resonator degrees of freedom to obtain an expression for the qubit matrix elements:
\begin{multline}
  \label{eq:qubit-dm-coherent-ics}
  \rho_\text{Q}^{(i)}(t) = \frac{Z^{(i)}(t)}{1 - 2H^{(i)}_{12}(t)}\exp\bigg[-|\alpha|^2\bigg(1 - 2H^{(i)}_{34}(t)\\
  -\frac{4H^{(i)}_{13}(t)H^{(i)}_{24}(t)}{1 - 2H^{(i)}_{12}(t)}\bigg)\bigg]\,.
\end{multline}
Taking $i=A,D$ and using the correct expressions for $H_{lk}^{(i)}(t)$ and $Z^{(i)}(t)$ from Eqs.~(\ref{eq:H(t)-matrix}) and (\ref{eq:H(t)-matrix-B}), we recover the qubit populations $\rho^{(A)}_\text{Q}(t) = q_{11},\rho^{(D)}_\text{Q}(t) =q_{22}$, as it should.

Again the analytics and numerics can be compared, using the same parameters as in the case of an initial thermal state. The result is shown in the right panel of Fig.~\ref{fig:coherence-evol} and we can see an excellent match between the two approaches. Also, if we take the zero temperature limit of Eq.~(\ref{eq:qubit-dm-coherent-ics}) with $\bar{n}\to 0$, we obtain the same expressions as in Ref.~\cite{PeixotodeFaria1999}.


\subsection{Open quantum Rabi model}
\label{sec:rabiRes}

Next, we look for a solution of the quantum Rabi model. The starting point for this discussion are the PDEs (\ref{eq:Rabi-DE-A}) to (\ref{eq:Rabi-DE-D}). In contrast to the previous case, the PDEs are coupled to each other so solving them is more challenging. Nonetheless, we have managed to obtain a perturbative analytical solution up to the second order in $g$. This perturbative expansion holds when $g$ is the smallest pertubative parameter such that
\begin{equation}
  \frac{g}{|\Delta|} \lessapprox 1\,.
\end{equation}

We begin by asserting that the full solutions to the PDEs (\ref{eq:Rabi-DE-A}) to (\ref{eq:Rabi-DE-D}) are of the form
\begin{multline}
  \label{eq:pert-theory-ansatz}
  A(t;z,z^*|w,v) = A^{(0)}(t;z,z^*|w,v) + gA^{(1)}(t;z,z^*|w,v) \\
  + g^2A^{(2)}(t;z,z^*|w,v) + \mathcal{O}(g^3)\,,
\end{multline}
and similarly for $B,C$ and $D$. Once we apply this expansion to the differential equations, we can group all the terms of same order in $g$ and solve the resulting equations separately.

The unperturbed propagators, proportional to $g^0$, are easily solved by noting that the quantum Rabi equations reduce to those of the dispersive Jaynes-Cummings model. Thus the unperturbed solutions are obtained straightforwardly from the results in the previous section, and they read:
\begin{subequations}
  \begin{align}
    A^{(0)}(t;z,z^*|w,v) &= q_{11}G(t; z,z^*|w,v)\,, \label{eq:A^(0)-sol} \\
    B^{(0)}(t;z,z^*|w,v) &= q_{12}\e^{-\i\omega t}G(t; z,z^*|w,v)\,, \label{eq:B^(0)-sol} \\
    C^{(0)}(t;z,z^*|w,v) &= q_{21}\e^{\i\omega t}G(t; z,z^*|w,v)\,, \label{eq:C^(0)-sol} \\
    D^{(0)}(t;z,z^*|w,v) &= q_{22}G(t; z,z^*|w,v)\,, \label{eq:D^(0)-sol}
  \end{align}
\end{subequations}
where we have introduced the Gaussian function
\begin{equation}
  \label{eq:Gaussian-part}
  G(t; z,z^*|w,v) = Z(t)\e^{2H_{12}(t)zz^* + 2H_{13}(t)zw + 2H_{24}(t)z^*v + 2H_{34}wv}\,,
\end{equation}
that is common for all the unperturbed propagators. The functions $H_{ij}$ and $Z(t)$ are the same as in Eq.~(\ref{eq:H(t)-matrix}) but with $g=0$.

\subsubsection{Perturbative solution: first order} \label{sec:pert-solut-open}

The details of the calculation for the first order correction $A^{(1)}(t;z,z^*|w,v)$ to the unperturbed propagator are given in Appendix \ref{sec:quantum-rabi-model-first-ord-details}. Here we just refer to the final result, which can be written as a linear polynomial in the complex variables with some time dependent coefficients $X^{(A)}_i$ as
\begin{equation}
  \label{eq:A^(1)-sol}
  A^{(1)} =
  \begin{aligned}[t]
    \i G\Big[&q_{12}\left(X_1^{(A)}z^* + X_2^{(A)}z + X_3^{(A)}w + X_4^{(A)}v\right) \\
    - &q_{21}\left(X_2^{(A)*}z^* + X_1^{(A)*}z + X_4^{(A)*}w + X_3^{(A)*}v\right)\Big]\,,
  \end{aligned}
\end{equation}
where $G$ is the Gaussian function of Eq.~(\ref{eq:Gaussian-part}). The coefficients $X^{(A)}_i$ depend on time and they are given in Appendix \ref{sec:quantum-rabi-model-first-ord-details}.

In a similar manner we can solve for the linear correction to the propagator $B^{(1)}$ as the governing PDE is almost the same as for $A^{(1)}$. The only difference between Eqs.~(\ref{eq:Rabi-DE-A}) and (\ref{eq:Rabi-DE-B}) is the appearance of the term $-\i\omega$. Otherwise the situation is equivalent, and we can write the solution as a linear polynomial
\begin{equation}
  \label{eq:B^(1)-sol}
  B^{(1)} = \i G\left[X_1^{(B)}z^* + X_2^{(B)}z + X_3^{(B)}w + X_4^{(B)}v\right]\,,
\end{equation}
where the coefficients are again time dependent and given in Appendix \ref{sec:quantum-rabi-model-first-ord-details}.

Similar methods can be used to derive the solutions for $C^{(1)}$ and $D^{(1)}$. Moreover, we can notice that the PDE for $C^{(1)}$, Eq.~(\ref{eq:Rabi-DE-C}), can be transformed into that of $B^{(1)}$ by replacing $-\i\omega\to\i\omega$ and $q_{11}\to q_{22}$ and $q_{22}\to q_{11}$. Similarly, the solution for $D^{(1)}$ can be obtained from that of $A^{(1)}$ by transforming  $-\i\omega\to\i\omega$, $q_{12}\to q_{21}$ and $q_{21}\to q_{12}$.

\paragraph{Dynamics with thermal initial condition}

We have now derived expressions for the propagators in the zeroth and first order in perturbation theory with respect to the qubit-boson coupling $g$. As in the case of the dispersive Jaynes-Cummings model, we did not specify the initial state of the boson when deriving the expressions for the propagator. Thus, the solution obtained above is completely general and can be applied to any initial separable state of the qubit-boson system.

In this subsection, we show explicitly the solution of the dynamics when the boson is initialized in a thermal state. To find the dynamics of the qubit-boson system we need to perform a Gaussian integration over the complex variables $w, v$, as shown in Eq.~(\ref{eq:rho-in-holomorphic_main}). We obtain the following expression:
\begin{widetext}
  \begin{equation}
    \label{eq:Rabi-solution-linear-in-g2}
    F_\text{B}\left(\rho(t); z, z^*\right) =
    \frac{\e^{\frac{\bar{n}}{\bar{n}+1}zz^*}}{1+\bar{n}}
    \begin{bmatrix}
      q_{11} & q_{12}\e^{-\i\omega t} \\
      q_{21}\e^{\i\omega t} & q_{22}
    \end{bmatrix} + g\left(\i \frac{z\e^{\frac{\bar{n}}{\bar{n}+1}zz^*}}{1+\bar{n}}
      \begin{bmatrix}
        \text{coeff}^{(A)}(t) & \text{coeff}^{(B)}(t) \\
        \text{coeff}^{(C)}(t) & \text{coeff}^{(D)}(t)
      \end{bmatrix}
      +\text{h.c.} \right) + \mathcal{O}(g^2)\,.
  \end{equation}
\end{widetext}
The first term is the zeroth order contribution, which becomes the exact solution for $g=0$, and the second term is the linear correction in $g$. Once again, the time dependent coefficients in the matrix are given in Appendix \ref{sec:quantum-rabi-model-first-ord-details}.

Now we can move from the Bargmann space back to the Fock space. This process is straightforward by using the holomorphic representation of the thermal state (see Eq.~(\ref{eq:thermal-state-holo})) and the relations (\ref{eq:a_L-map}) to (\ref{eq:a_R^dag-map}) for the map between the complex variables to the ladder operators. We finally obtain
\begin{multline}
  \label{eq:Rabi-solution-linear-Fock-space}
  \rho(t) =
      \begin{bmatrix}
      q_{11} & q_{12}\e^{-\i\omega t} \\
      q_{21}\e^{\i\omega t} & q_{22}
      \end{bmatrix}
      \otimes\rho_\text{Th} \\
      + g\left(\i 
    \begin{bmatrix}
      \text{coeff}^{(A)}(t) & \text{coeff}^{(B)}(t) \\
      \text{coeff}^{(C)}(t) & \text{coeff}^{(D)}(t)
    \end{bmatrix}
    \otimes a^\dagger\rho_\text{Th} +\text{h.c.} \right)\mar{+ \mathcal{O}(g^2)} \,.
\end{multline}

From this expression it is immediately clear that the perturbative solution in the first order of $g$ does not bring any contribution to the dynamics of many interesting observables. For instance, any local qubit observable is not modified by the term proportional to $g$, because  computing the expectation values of any qubit observable requires taking the trace over the boson Hilbert space, and $\Tr[a^\dagger\rho_\text{Th}] = 0$. Therefore, only the observables that are odd in the number of ladder operators get a non-zero contribution from the first order of Eq.~\eqref{eq:Rabi-solution-linear-Fock-space}.

\begin{figure}
  \centering
  \includegraphics[width=\linewidth]{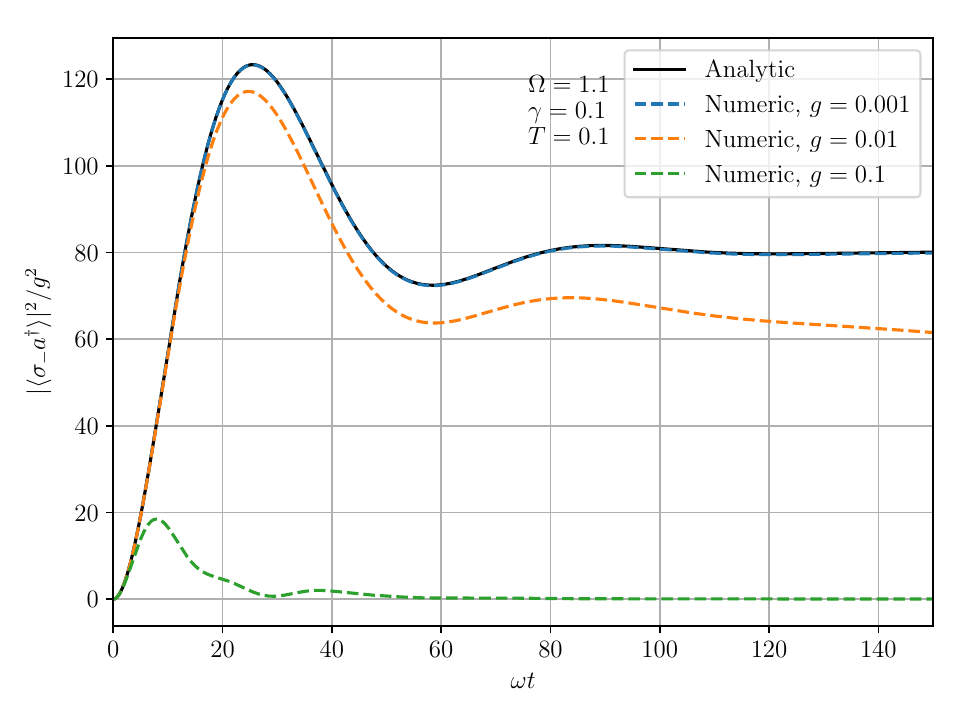}
  \caption{$|\langle \sigma_- a^\dagger\rangle|^2/g^2$ as a function of normalized time for several values of the qubit-boson coupling $g$ with other parameters set to $\omega = 1,\ \Omega = 1.1$ and $\gamma = T = 0.1$. The initial state of the qubit is $\ket{e}$. The numerical results (dashed lines) agree with the analytical solution (black solid line) from first order of perturbation theory for longer times as $g$ gets smaller.}
  \label{fig:Rabi-1st-ord}
\end{figure}

In order to test the validity of the analytical solution, we compute the expectation value $\langle \sigma_-a^\dagger\rangle$, which is odd in the number of ladder operators and captures the presence of coherences of the form $\ket{0,n}\bra{1,n-1}$. We compare this result to the numerics for different values of $g$ and the results are depicted in Fig.~\ref{fig:Rabi-1st-ord}. We see that the period of time during which the numerical and analytical solutions agree depends on the magnitude of $g$, as expected. 

Note that, as discussed in Sec.~\ref{sec:steadyState}, the steady state of the master equation for the quantum Rabi model is always unique. In contrast, the steady state solution of Eq.~\eqref{eq:Rabi-solution-linear-Fock-space} depends on the initial conditions of the qubit, as shown in Eqs.~\eqref{eq:F_B11-corr_coeffA} to~\eqref{eq:F_B22-corr_coeffD}. This means that at the first order of the perturbation theory we still find a 2-dimensional family of steady states. As a consequence, the first-order perturbative solution is valid only for early times, while it must give wrong predictions for late times and for the steady state.

\begin{figure}[h]
  \centering
  \includegraphics[width=\linewidth]{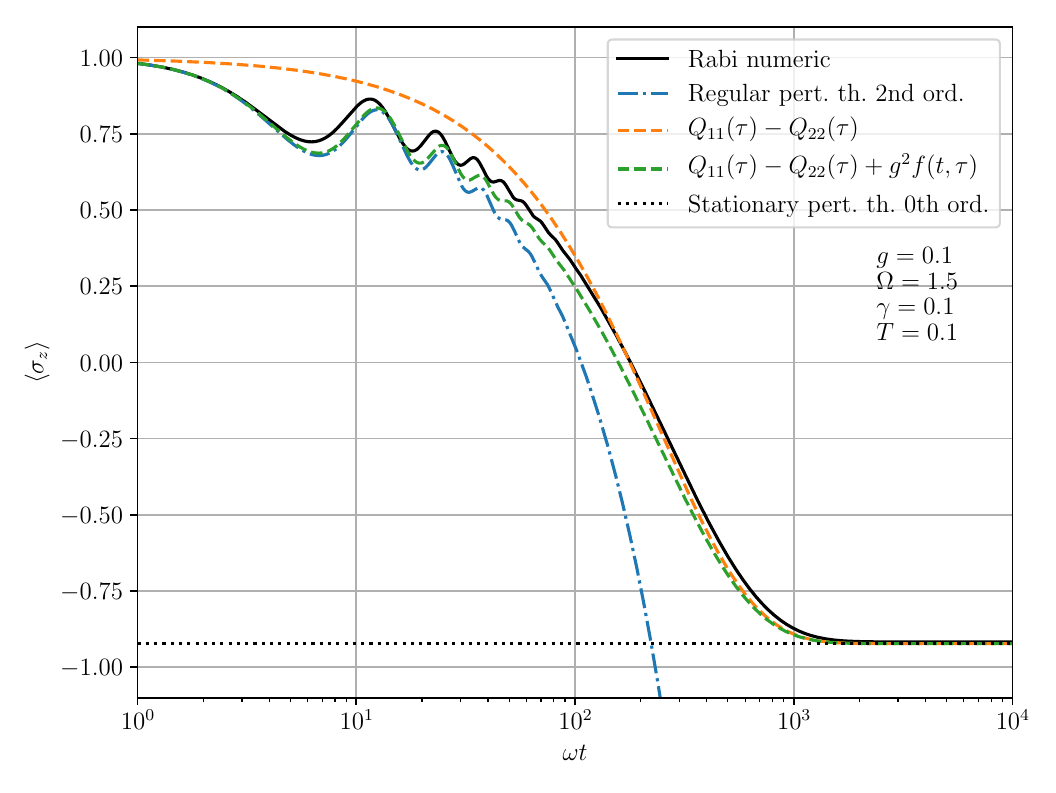}
  \caption{\mar{Numerically exact} (black solid line) and analytical solutions for $\langle\sigma_z\rangle$ in the open quantum Rabi model with simulation parameters $\omega = 1,\ \Omega = 1.5$ and $g = \gamma = T = 0.1$. The qubit is initialized in the excited state. The direct application of the second order perturbation theory (blue dash-dotted line) leads to accurate results for small times, but diverges from the correct dynamics at later times. \ant{This divergence can be controlled by applying multiscale perturbation theory \mar{that introduces a long time scale $\tau = g^2t$. The difference between the qubit populations $Q_{11}(\tau)-Q_{22}(\tau)$ obtained from multiscaling (orange dashed line)} improves the prediction at late times but fails to capture the behavior at early times. \mar{The second-order expression coming from multiscaling, Eq.~\eqref{eq:2nd-ord-ms-expsigmaz} (dashed green line)}}, improves the result for early times, while still converging to the steady state result of Eq.~(\ref{eq:sigmaz-steadyRabi}) (dotted black line).  \mar{Note that the exact numerical results agree very well with the perturbative analytical solution, even for relatively large qubit-boson coupling values, such as $g=0.1$, which fall within the so-called ``perturbative region of the ultra-strong coupling regime" \cite{Forn-Diaz2019}.}}
  \label{fig:Rabi-2nd-ord-sigmaz}
\end{figure}

\subsubsection{Pertubative solution: second order with multiscaling} \label{sec:pert-solut-quant}

To find the small-$g$ contribution to the dynamics of a broader class of observables, including the expectation values of local Pauli operators of the qubit, we apply second order perturbation theory. The direct application of regular perturbation theory, i.e. computing the correction $A^{(2)}(t; z,z^*|w,v)$ from Eq.~(\ref{eq:pert-theory-ansatz}), gives rise to so-called secular terms, which are linear in time and thus divergent as $t\to\infty$. This issue can be cured by using renormalized perturbation theory that introduces multiple timescales. Multiscale perturbation theory is explained, for example, in chapter 4 of \cite{Kevorkian1996}, chapter 11 of \cite{Verhulst2006} and chapter 7 of \cite{Strogatz2018}.

For simplicity, we do not compute the second order corrections at the level of propagators. Instead, we directly assume that the initial state of the boson is thermal. Thus we solve for the elements $\left[F_\text{B}\left(\rho(t); z,z^*\right)\right]_{ij} \equiv f_{ij}$ of the Bargmann space density matrix. We can express them as:
\begin{equation}
    \label{eqn:pertExpan_f}
    f_{ij} = f_{ij}^{(0)} + g f_{ij}^{(1)} + g^2 f_{ij}^{(2)} + \mathcal{O}(g^3).
\end{equation}
$f_{ij}^{(0)}$ and $f_{ij}^{(1)}$ can be immediately obtained from the first-order solution in Eq.~\eqref{eq:Rabi-solution-linear-in-g2}.
Moreover, it can be shown that their second-order correction from standard perturbation theory can be expressed as:
\begin{multline}
  \label{eq:Rabi-2nd-ord-sol}
  f_{ij}^{(2)}(t;z,z^*) = \Big[x^{(2,0)}_{ij}(t)z^2 + x^{(1,1)}_{ij}(t)zz^* \\
  + x^{(0,2)}_{ij}(t)z^{*2} + x^{(0,0)}_{ij}(t)\Big]\e^{\frac{\bar{n}}{1+\bar{n}}zz^*}\,,
\end{multline}
where the time dependent coefficients $x_{ij}^{(a,b)}$ are given in the Appendix \ref{sec:quantum-rabi-model-sec-ord} together with further details on the derivation of Eq.~\eqref{eq:Rabi-2nd-ord-sol}.

\begin{figure}
  \centering
  \includegraphics[width=\linewidth]{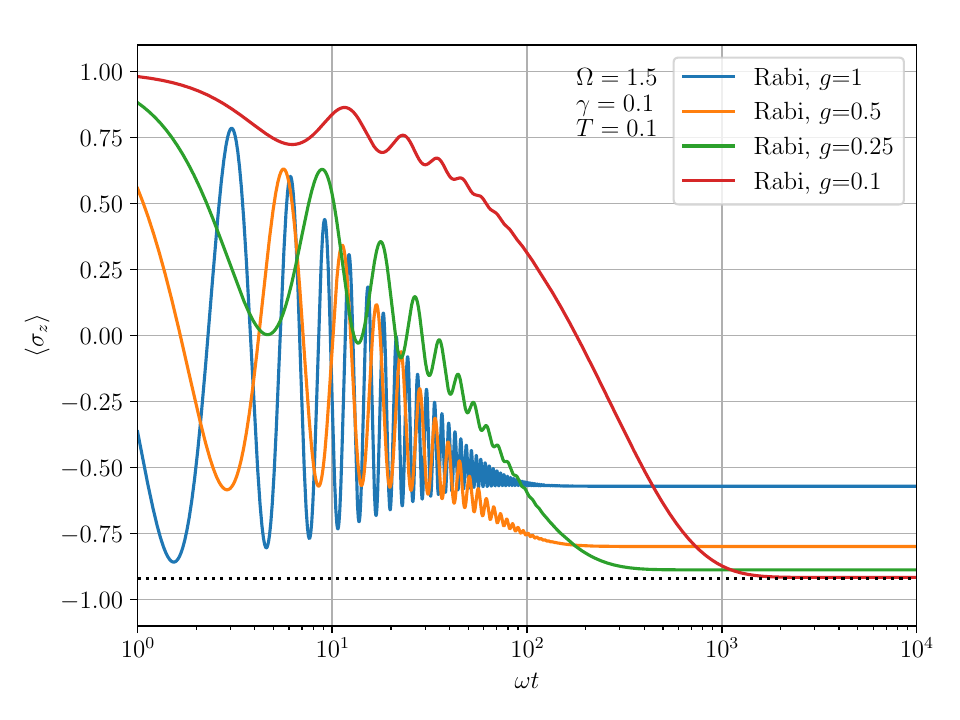}
  \caption{Numerical results for $\langle\sigma_z\rangle$ in the open quantum Rabi model with varying qubit--boson couplings $g$. The other parameters are fixed as $\omega=1,\ \Omega = 1.5$ and $\gamma = T = 0.1$. The qubit is initialized in the excited state. The numerical solutions converge towards the analytical asymptotic value predicted by Eq.~(\ref{eq:sigmaz-steadyRabi}) (dotted black line) for smaller values of $g$. }
  \label{fig:QRabi_szExp_varyg}
\end{figure}

\begin{figure*}
  \centering
  \begin{minipage}{0.5\textwidth}
    \centering
    \includegraphics[width=\linewidth]{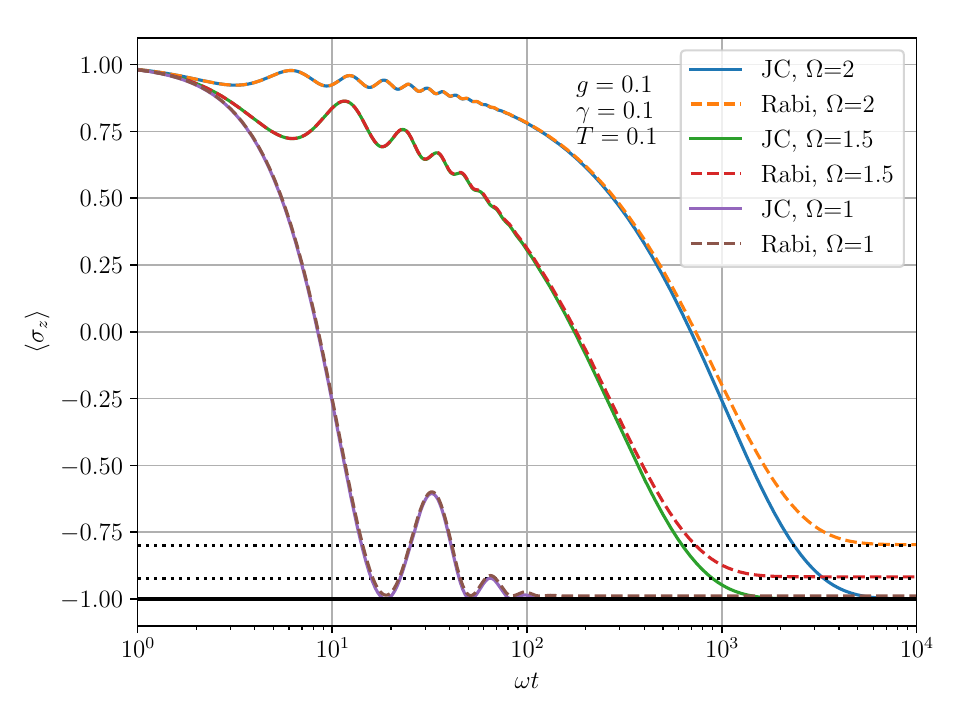}
  \end{minipage}%
  \begin{minipage}{0.5\textwidth}
    \centering
    \includegraphics[width=\linewidth]{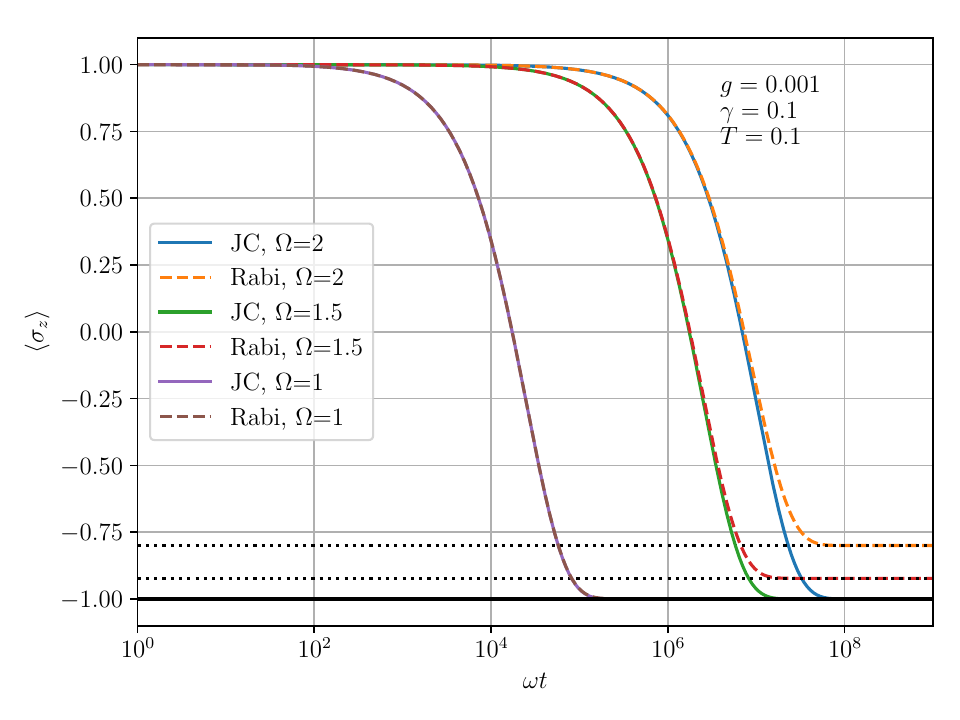}
  \end{minipage}
  \caption{Numerical results for $\langle\sigma_z\rangle$ in the open quantum Rabi and Jaynes-Cummings models with varying boson frequency $\Omega$ \ant{for two different choices of $g$. On the left $g=0.1$ and on the right $g=0.001$.} The other parameters are fixed as $\omega=1$ and $\gamma = T = 0.1$.  The qubit is initialized in the excited state.  The solid lines represent the dynamics of the Jaynes-Cummings model, while the dashed lines illustrate that of the quantum Rabi model.  The numerical solutions converge towards the analytical asymptotic values predicted by Eqs.~(\ref{eq:sigmaz-steadyRabi}) in the case of Rabi (dotted black  lines) and~\eqref{eq:sigmaz-steadyJC} in the case of Jaynes-Cummings (solid black  line).  Note that for early times the solutions of both models coincide due to the small value of $g$. Moreover, note that for $g/\Delta\ll 1$, with $\Delta=\Omega-\omega$, we are in the dispersive regime of the Jaynes-Cummings model. However, the qubit state relaxes towards the stationary analytical prediction in a much longer timescale than for $\Delta =0$, due to the small remainder in the approximated Hamiltonian of Eq.~\eqref{eqn:dispersiveHam}. \mar{Once again, we observe good agreement between the perturbative analytical solution and the exact numerical results also for $g=0.1$, approaching the perturbative region of the ultra-strong coupling regime \cite{Forn-Diaz2019}.}}
  \label{fig:Rabi_JC_varyingOmega}
\end{figure*}
Here the coefficients $x_{ij}^{(0,0)}$ include the problematic secular terms, making the solution valid only for small times, and motivating the use of multiscale perturbation theory. \mar{Multiscale perturbation theory treats separately the two timescales of the problem, namely the one driven by $t$ and the one by $\tau=g^2 t$.} Through this process we can obtain some expressions $Q_{ij}(\tau)$ for the late-time evolution of the density matrix elements of the qubit that \ant{fix the time divergence at second order}. In other words, through multiscale perturbation theory we obtain a \ant{second}-order solution for $f_{ij}^{(2)}$ that is a bounded function of time. This function relaxes to a physical steady state, being accurate at \mar{zeroth} order, and not depending on the initial conditions. The formulae for the exact expressions for $Q_{ij}(\tau)$ are given in Appendix \ref{sec:quantum-rabi-model-sec-ord}. 

We compare the results obtained from second order corrections to numerics by computing the evolution of $\langle\sigma_z\rangle$. In Fig.~\ref{fig:Rabi-2nd-ord-sigmaz} we can see how at small times the second order solution from Eq.~(\ref{eq:Rabi-2nd-ord-sol}), obtained via the application of regular perturbation theory, works \ant{as well as} the \ant{second}-order one obtained from multiscaling. However, the solution from Eq.~(\ref{eq:Rabi-2nd-ord-sol}) diverges quickly from the correct numerical solution as time grows due to the secular terms. At sufficiently large times the best analytical result is given by the corrected expression \mar{through multiscaling}, which is
\ant{
  \begin{multline}
    \label{eq:2nd-ord-ms-expsigmaz}
    \langle\sigma_z\rangle \approx Q_{11}(\tau) - Q_{22}(\tau) + g^2\Big[\tilde{x}_{11}^{(0,0)}(t,\tau) - \tilde{x}_{22}^{(0,0)}(t,\tau)\\
    + (1+\bar{n})\left(\tilde{x}_{11}^{(1,1)}(t,\tau) - \tilde{x}_{22}^{(1,1)}(t,\tau)\right)\Big]\,,
  \end{multline}
  where $\tilde{x}_{ij}^{(a,b)}(t,\tau)$ are the coefficient functions of Eq.~(\ref{eq:Rabi-2nd-ord-sol}) but with multiscaling taken into account when solving for them. \mar{As usual in multiscale perturbation theory,} we cannot rule out the presence of other terms of order $g^2$ that might arise from solvability conditions involved in higher orders of the multiscale perturbation theory.  \mar{That said, we note that the terms going as $g^2$ in Eq.~\eqref{eq:2nd-ord-ms-expsigmaz} may considerably enhance the precision of the perturbative solution also at intermediate times, as shown through an example in Fig.~\ref{fig:Rabi-2nd-ord-sigmaz}.}
}

Note that the limit $t\rightarrow\infty$ of the corrected \ant{second} order result for $\langle\sigma_z\rangle$ does not depend on the initial conditions \ant{of the qubit}, as shown in Appendix~\ref{sec:quantum-rabi-model-sec-ord}, in contrast to the late-time limit of Eq.~(\ref{eq:Rabi-solution-linear-Fock-space}), corresponding to the uncorrected zeroth order result of $\langle\sigma_z\rangle = q_{11} - q_{22}$. This convergence is consistent with the fact that there is a unique steady state of the master equation, as discussed in Sec.~\ref{sec:steadyState}. We see that $\langle\sigma_z\rangle$ converges to a stationary value that can be computed by using the regular perturbation theory at second order in the static case, which is discussed next. 

\subsubsection{Zeroth-order perturbative solution for the steady state}
\label{sec:steady}

We can obtain a perturbative solution for the steady state of the local master equation for the quantum Rabi model by setting the time derivative to zero in Eqs.~\eqref{eq:Rabi-DE-A} to~\eqref{eq:Rabi-DE-D}. 
As explained in Sec.~\ref{sec:steadyState}, if $g=0$ there is a whole family of reduced steady states of the qubit, defined in Eq.~\eqref{eqn:ssSeparable}. This means that, at the zeroth order of static perturbation theory, there is a freedom in the steady state solution for $f_{ij}^{(0)}$ in Eq.~\eqref{eqn:pertExpan_f}, which can be written as:
\begin{equation}
    \label{eqn:f_ij0}
    f_{ij}^{(0)} = Z_{ij}^{(ss)} e^{\frac{\bar{n}}{1+\bar{n}}z z^*},
\end{equation}
where $Z_{12}^{(ss)}=Z_{21}^{(ss)}=0$ (no steady-state coherences according to Eq.~\eqref{eqn:ssSeparable}), while $Z_{11}^{(ss)}$ and $Z_{22}^{(ss)}$ are not fixed a priori, but depend on the initial conditions of the qubit. Note that the steady state of the boson is fixed (thermal Gibbs state as explained in Eq.~\eqref{eq:thermal-state-holo}), while the state of the qubit is not. This freedom is not removed by first-order perturbation theory, as explained after Eq.~\eqref{eq:Rabi-solution-linear-Fock-space}. 

Next, if we solve for the static second-order equations, we obtain a set of equations for the coefficients of $f_{ij}^{(2)}$ in Eq.~\eqref{eq:Rabi-2nd-ord-sol} that in turn depend on $Z_{ii}^{(ss)}$. This set of equations fixes the value of $Z_{ii}^{(ss)}$, and hence of $f_{ii}^{(0)}$ in the steady state. In other words, through second-order perturbation theory we obtain a \textit{consistency condition} for the zeroth-order steady state solution, breaking the freedom in the choice of the steady state of the qubit for $g=0$. 

Focusing on the qubit only, this consistency condition implies:
\begin{equation}
  \label{eq:sigmaz-steadyRabi}
  \begin{split}
  &\lim_{t\to\infty}\langle\sigma_z\rangle^\text{(Rabi)} = -\frac{8\omega\Omega}{(1 + 2\bar{n})(\gamma^2 + 4(\omega^2 + \Omega^2))}\mar{+ \mathcal{O}(g^2)}\,.\\
  \end{split}
\end{equation}
This solution is valid for the Rabi model only. In fact, we can perform the same calculations also for the non-dispersive Jaynes-Cummings model; note this is the only case in this work where we consider Eq.~\eqref{eqn:masterEqSimple} with $\alpha=\text{JC}$. Interestingly, the steady-state solution is different:
\begin{equation}
  \label{eq:sigmaz-steadyJC}
  \begin{split}
  &\lim_{t\to\infty}\langle\sigma_z\rangle^\text{(JC)} = -\frac{1}{1 + 2\bar{n}}\mar{+ \mathcal{O}(g^2)}\,.\\
  \end{split}
\end{equation}

We observe that the zeroth-order steady state populations of the Jaynes-Cummings model are given by the standard Gibbs thermal state $\exp(-\beta \omega \sigma_z/2)$. In contrast, the expression for the stationary populations of the Rabi model is more convoluted, and depends non-trivially on the qubit and boson frequencies. It converges to the value predicted by the Gibbs state only for $\omega=\Omega$, $\gamma\ll\omega$. We thus obtain a non-perturbative difference between the late-time predictions of the Jaynes-Cummings and Rabi model, which is independent of the value of the qubit-boson coupling $g$. Note that Eqs.~\eqref{eq:sigmaz-steadyRabi} and~\eqref{eq:sigmaz-steadyJC} are valid also at resonance ($\Delta =0$).

\ant{
  Eq.~(\ref{eq:sigmaz-steadyRabi}) can also be derived from the dynamical case, by taking the inifinite time limit of Eq.~(\ref{eq:2nd-ord-ms-expsigmaz}). This will yield
  \begin{widetext}
\begin{equation}
  \lim_{t\to\infty}\langle\sigma_z\rangle^\text{(Rabi)} \mar{\approx} -\frac{8\omega\Omega}{(1+2\bar{n})\left(\gamma^2+4\left(\omega^2+\Omega^2\right)\right)} + g^2\frac{256\gamma^2\omega\Omega}{\left(\gamma^2+4\left(\omega^2+\Omega^2\right)\right) \left(\gamma^4 + 8\gamma^2\left(\omega^2+\Omega^2\right) + 16\left(\omega^2-\Omega^2\right)^2\right)}\,,
\end{equation}
\end{widetext}
where the first term is the zeroth order steady state solution $Q_{11}(g^2t) - Q_{22}(g^2t)$ and the second corresponds to the second order correction.
}

We check the validity of Eqs.~\eqref{eq:sigmaz-steadyRabi} and~\eqref{eq:sigmaz-steadyJC} by comparing the late-time results of non-perturbative numerical simulations of these models with the zeroth-order  predictions for their stationary populations. The results are shown in Figs.~\ref{fig:QRabi_szExp_varyg} and~\ref{fig:Rabi_JC_varyingOmega}, where we plot different solutions by varying $g$ and $\Omega$ respectively. All the numerical solutions converge to the analytical stationary values. In particular, note that in Fig.~\ref{fig:Rabi_JC_varyingOmega} all the Jaynes-Cummings evolutions converge to Eq.~\eqref{eq:sigmaz-steadyJC} independently of $\Omega$, while the Rabi solutions are sensitive to $\Omega$ and converge to the values predicted by Eq.~\eqref{eq:sigmaz-steadyRabi}. \mar{Moreover, from Fig.~\ref{fig:Rabi_JC_varyingOmega} we observe that decreasing $g$ increases the thermalization time of both models. For $g=0$, the qubit does not relaxe and its steady state is not unique.}

\section{Conclusions} \label{sec:conclusions}

In this work, we have leveraged the holomorphic formalism in Bargmann space to obtain new analytical solutions of the open \ant{dispersive} Jaynes-Cummings and quantum Rabi models. Our analysis focused on open systems governed by local Lindblad equations, where a thermal bath acts on the bosonic mode only, and the qubit-boson coupling $g$ is weak. We have  addressed  separately the dispersive Jaynes-Cummings model and the quantum Rabi model. We have not considered the dynamics of the non-dispersive Jaynes-Cummings model as a separate scenario. Nonetheless, we have analyzed the steady-state equations for the non-dispersive Jaynes-Cummings model independently, demonstrating that their solutions can deviate from those of the quantum Rabi model.
 \mar{The dynamical perturbative solution for the non-dispersive Jaynes-Cummings model may be obtained quite straightforwardly by following the same lines of our solution for the quantum Rabi model. Indeed, the differential equations of these two models are similar, with those of the non-dispersive Jaynes-Cummings model being slightly simpler due to the asbsence of the counter-rotating terms. Moreover, for small qubit-boson coupling $g$, the Jaynes-Cummings and quantum Rabi models are usually considered equivalent \ant{\cite{Forn-Diaz2019,DeBernardis2024,PhysRevA.86.033837}}. We thus expect their dynamical solution to be equivalent, at least at early times.}

For the dispersive Jaynes-Cummings model, we have obtained an exact solution presented in its full generality in Sec.~\ref{sec:openDis}. The only assumption we have made is an initial separable state of the qubit-boson system, thus extending the generality of previous results that relied on a fixed thermal or coherent initial state of the boson. We have validated our solution by checking that it indeed recovers the results available in the literature under the suitable choice of initial state of the boson, and through numerical simulations shown in Fig.~\ref{fig:coherence-evol}. 

In the case of the quantum Rabi model, 
we have derived a perturbative solution up to the second order in $g$, which is valid away from resonances. The solution is presented in Sec.~\ref{sec:rabiRes}. The first-order solution is general and requires only the assumption of a separable initial state for the system, but it remains valid solely at early times. In contrast, the second-order solution assumes a thermal initial state for the bosonic mode and its validity can be extended to late times through multiscale perturbation theory.  To the best of our knowledge, no prior analytical solution for the open quantum Rabi model at finite temperature has been reported. Our results thus provide a novel analytical benchmark for this class of systems.

Finally, by leveraging second-order perturbation theory we have also obtained the zeroth-order solution for the steady state of the quantum Rabi model and non-dispersive Jaynes-Cummings model, presented in Sec.~\ref{sec:steady}. Clearly, the stationary state of the bosonic mode is thermal at the zeroth order in $g$. In contrast, the steady state of the qubit depends on the model we choose: it is thermal for the Jaynes-Cummings model, while it is diagonal but non-thermal for the quantum Rabi model. In the latter scenario, the stationary populations depend on both the qubit and boson frequencies. We thus found a non-perturbative difference in the late-time dynamics of these two models, which are usually considered equivalent if $g$ is small. We have validated our analytical results through exact numerical simulations, obtaining perfect agreement between the two approaches for weak $g$. 

Our findings are valuable for the study of open qubit-boson systems in the limit of small inter-system coupling $g$, and offer new analytical insights into the open dispersive Jaynes-Cummings and quantum Rabi models. Furthermore, the versatility and generality of the holomorphic formalism demonstrated in this work can be leveraged to extend our results to unexplored scenarios, including higher-order perturbative solutions, multi-qubit systems, systems with additional dissipation channels acting on the qubit, or regimes beyond the weak-coupling limit of $g$.

\begin{acknowledgments}
The authors thank Masaaki Tokieda for pointing out the solvability of the open dispersive Jaynes-Cummings model. We acknowledge the financial support of the Research Council of Finland through the Centre of Excellence program grant 336810 and through the Finnish Quantum Flagship project 358878 (UH) and 358877  (Aalto). M.C. acknowledges funding from COQUSY project PID2022-140506NB-C21 funded by MCIN/AEI/10.13039/501100011033.
\end{acknowledgments}

\appendix

\section{Passing from the Fock space to the holomorphic formalism in Bargmann space}
\label{app:Holo}

In this appendix we give a self-consistent introduction to the holomorphic formalism for the quantum harmonic oscillator. Moreover, we present all the necessary tools to derive the holomorphic (Bargmann) representation of Lindblad master equations.
The discussion follows the one in Ref.~\cite{Vourdas1994}.

\subsection{Holomorphic representation of states and operators} \label{sec:holom-repr-stat}

The quantum optical coherent states $\ket{z}$ are constructed as eigenstates of the annihilation operator through $a\ket{z} = z\ket{z}$ with $z\in \mathbb{C}$. The solution to this eigenvalue problem is given in the Fock space as \cite{Glauber1963}
\begin{equation}
  \label{eq:coherent-state-def-Fock}
  \ket{z} = \e^{-\frac{|z|^2}{2}}\sum_{n=0}^\infty\frac{z^n}{\sqrt{n!}}\ket{n}\,,
\end{equation}
where $\ket{n}$ is the basis of excitation number in the Fock space. An important property of the coherent states is the resolution of the identity as
\begin{equation}
  \label{eq:identity-op-coherent-states}
  \mathbb{1} = \frac{1}{\pi}\int\d^2z\ket{z}\bra{z} = \frac{1}{2\pi}\int\d z\d z^*\ket{z}\bra{z}\,,
\end{equation}
where $z^*$ is the \textit{formal} complex conjugate of $z$ and the integration is over the entire complex plane. The word \textit{formal} means that although $z$ and $z^*$ can be taken as complex conjugates of each other, they are independent integration variables in the complex plane.\footnote{This is due to how differentiation is defined in the complex plane via the Wirtinger derivative. One can show that indeed $\partial z/\partial z^* = \partial z^*/\partial z = 0$} Therefore the notation $\d z\d z^*$ corresponds to integration over the real 2D plane, embedded in $\mathbb{C}^2$ \cite{ZinnJustin2005}.

Let us now consider an arbitrary normalized state in the Fock space defined as
\begin{equation}
  \label{eq:state-in-Fock}
  \ket{\phi} = \sum_{n=0}^\infty\phi_n\ket{n}\,, \hspace{0.5cm} \sum_{n=0}^\infty|\phi_n|^2 = 1\,.
\end{equation}
In what follows, we will often also encounter conjugated bra-states, which are defined as $\bra{\phi^*} = \left(\ket{\phi^*}\right)^\dagger = \sum_n\phi_n\bra{n}$.

In the holomorphic representation the arbitrary state in equation (\ref{eq:state-in-Fock}) is given as an analytic function $f_\text{B}(\ket{\phi}; z)$ of the complex variables $z$ (the subscript B refers to Bargmann), which is obtained by taking the inner product of the state $\ket{\phi}$ in Eq.~(\ref{eq:state-in-Fock}) with a coherent state $\ket{z}$ in Eq.~(\ref{eq:coherent-state-def-Fock}):
\begin{equation}
  \label{eq:state-map-Fock-to-Bargmann}
  f_\text{B}(\ket{\phi}; z) \equiv \e^{\frac{|z| 2}{2}}\braket{z^*|\phi} = \sum_{n=0}^\infty\phi_n\frac{z^n}{\sqrt{n!}}\,.
\end{equation}
We notice that the Fock space state $\ket{\phi}$ transforms into a polynomial in $z$, which is a holomorphic function in $\mathbb{C}$.

The inner product in the holomorphic representation can be obtained by using the resolution of the identity of the coherent states in equation (\ref{eq:identity-op-coherent-states}) in the middle of the Fock space inner product as
\begin{equation}
  \label{eq:inner-product-holomorphic}
  \begin{aligned}
    \braket{\psi^*|\phi} &= \frac{1}{\pi}\int\d^2z \braket{\psi^*|z}\braket{z|\phi} \\
    &= \frac{1}{\pi}\int\d^2z \e^{-|z|^2}f_\text{B}\left(\ket{\psi}; z\right)f_\text{B}\left(\ket{\phi}; z^*\right)\,.
  \end{aligned}
\end{equation}

The holomorphic representation for the operators arises in a similar manner as the one for quantum states. We construct the holomorphic function $F_\text{B}\left(O; z, z^*\right)$ for the operator $O$ from the matrix element $\braket{z^*|O|z^*}$ of the operator with respect to the coherent states:
\begin{equation}
  \label{eq:operator-map-to-Bargmann}
 F_\text{B}\left(O; z, z^*\right) \equiv \e^{|z|^2}\braket{z^*|O|z^*} = \sum_{n,m=0}^\infty O_{nm}\frac{z^nz^{*m}}{\sqrt{n!}\sqrt{m!}}\,.
\end{equation}


The product of operators can be shown to have the following representation in the Bargmann space:
\begin{multline}
  \label{eq:operator-product-Bargmann}
  F_\text{B}\left(O_1O_2; z, z^*\right) = \frac{1}{\pi}\int\d^2w \e^{-|w|^2}F_\text{B}\left(O_1; z, w\right)\\
  \times F_\text{B}\left(O_2; w^*, z^*\right) \,.
\end{multline}
The tensor product of operators $O_1O_2 \equiv O_1\otimes O_2$ separates the function for the holomorphic representation as
\begin{equation}
  F_\text{B}\left(O_1O_2; z, z^*| w, w^*\right) = F_\text{B}\left(O_1; z, z^*\right)F_\text{B}\left(O_2; w, w^*\right)\,.
\end{equation}

The trace of an operator can be obtained from its holomorphic representation by integrating out the complex variables over the whole space
\begin{equation}
  \label{eq:trace-in-Bargmann}
  \Tr[O] = \frac{1}{\pi}\int\d^2z \e^{-|z|^2}F_\text{B}\left(O; z, z^*\right)\,.
\end{equation}
This can be shown to hold true by using the following important identity for the Gaussian moments in the complex plane
\begin{equation}
  \label{eq:Gaussian-moments-integral}
  \int \d^2z\e^{-|z|^2}z^nz^{*m} = \pi n!\delta_{nm}\,,
\end{equation}
which can be proven using Wick's theorem \cite{ZinnJustin2005}.

\subsection{Holomorphic representation of linear maps on operators}\label{sec:holom-repr-super}

It is well known that a linear map on a matrix space can always be represented as matrix on the vector space defined by reshaping the matrix space (see e.g. \cite{Gyamfi2020} or chapter~4 of \cite{HoJo1991}). By reshaping we mean using the one-to-one correspondence between outer and tensor product
$$\ket{a}\bra{b} \to \ket{a}\otimes\ket{b^*} \equiv \kket{a,b}$$
and writing matrices as column vectors obtained by stacking matrix rows one above the other.
Correspondingly, we vectorize an operator in the Fock space as
\begin{equation}
  \label{eq:vectorized-O}
  O = \sum_{n,m=0}^\infty O_{nm}\ket{n}\bra{m} \to \kket{O} = \sum_{n,m=0}^\infty O_{nm}\ket{n}\otimes\ket{m}\,.
\end{equation}
We can see that this process transforms a matrix in basis $\ket{n}\bra{m}$ to a vector in basis $\ket{n}\otimes\ket{m}$.


We write linear maps $S[O]$ of operators onto operators as
\begin{equation}
  \label{eq:superoperator-in-Lioville-space}
  S = \sum_{i,j,k,l}S_{ij,kl}\kket{i,j}\bbra{k,l}\,,
\end{equation}
where $S_{ij, kl} = \bbra{i,j}S\kket{k,l}$ is the matrix element of the map $S$ in the reshaped vector space. Then we define the Bargmann space representation of the linear map as
\begin{equation}
  \label{eq:superoperator-mapping-to-holomorphic}
  \begin{aligned}
    \mathcal{F}_\text{B}\left(S; z, z^*|w, w^*\right) &\equiv \e^{|z|^2 + |w|^2}\bbra{z^*, z^*}S\kket{w, w} \\
                                                      &= \sum_{i,j,k,l}S_{ij,kl}\frac{z^iz^{*j}w^kw^{*l}}{\sqrt{i!}\sqrt{j!}\sqrt{k!}\sqrt{l!}}\,.
  \end{aligned}
\end{equation}
It is noteworthy that for describing the linear maps on operators in the Bargmann space, we need four complex variables $z,z^*,w$ and $w^*$. The reason for this becomes clear when we think about $\mathcal{F}_\text{B}(S; z,z^*|w,w^*)$ as an integration kernel, which tranforms Bargmann space operators $F_\text{B}(O;z,z^*)$ to other operators via integration. This transformation is written as
\begin{multline}
  \label{eq:operator-transformation-under-superoperator-holomorphic}
  F_\text{B}\left(S[O]; z, z^*\right) = \frac{1}{\pi^2}\int\d^2w_1\d^2w_2\e^{-|w_1|^2 - |w_2|^2}\\
  \times\mathcal{F}_\text{B}\left(S; z, z^*|w_1, w_2^*\right)F_\text{B}\left(O; w_1^*, w_2\right)\,.
\end{multline}
Now the other two complex variables are integrated out such that we are left with something that is of the operator form, depending only of $z$ and $z^*$. A concrete example of this was given in Eq.~(\ref{eq:time-evolved-rho-holo}) for the density matrix.

\section{Quantum Rabi model: first order calculation details} \label{sec:quantum-rabi-model-first-ord-details}
In this appendix we present the details for the discussion of section \ref{sec:pert-solut-open}, where the first order perturbative solution of the open quantum Rabi model was considered. 

We begin from the linear correction to the propagator $A(t;z,z^*|w,v)$ (see Eq.~(\ref{eq:pert-theory-ansatz})), which satisfies the following differential equation (obtained from Eq.~(\ref{eq:Rabi-DE-A}))
\begin{multline}
  \label{eq:linear-correction-Rabi}
  \dot{A}^{(1)}(t;z,z^*|w,v) = \mathcal{D}(z,z^*)A^{(1)}(t;z,z^*|w,v) \\
  + \mathcal{N}^{(A)}(t;z,z^*|w,v)\,,
\end{multline}
where $\mathcal{D}(z,z^*)$ is a differential operator defined in Eq.~(\ref{eq:diff-op-DSecond}) and $\mathcal{N}^{(A)}(t;z,z^*|w,v)$ is a non-homogeneous term of the form
\begin{multline}
  \label{eq:non-homog-term-Rabi}
  \mathcal{N}^{(A)}(t;z,z^*|w,v) = \i\bigg[q_{12}\e^{-\i\omega t}\left(z^* + \frac{\partial}{\partial z^*}\right) \\
  - q_{21}\e^{\i\omega t}\left(z + \frac{\partial}{\partial z}\right)\bigg]G(t;z,z^*|w,v)\,,
\end{multline}
with $G(t;z,z^*|w,v)$ being the Gaussian defined in Eq.~(\ref{eq:Gaussian-part}).

Based on this, we are motivated to apply the following ansatz as the solution of the PDE in Eq.~(\ref{eq:linear-correction-Rabi}):
\begin{multline}
  \label{eq:A^(1)-ansatz}
  A^{(1)}(t; z,z^*|w,v) = \frac{1}{\pi^2}\int_0^t\d s \int\d^2\xi\d^2\zeta \e^{-|\xi|^2 - |\zeta|^2}\\
  \times G(t-s; z,z^*|\xi,\zeta^*)\mathcal{N}^{(A)}(s; \xi^*, \zeta|w,v)\,,
\end{multline}
where the Gaussian $G(t;z,z^*|w,v)$ acts as a Green's function of the homogeneous problem.

This integral can be computed exactly. After explicitly evaluating the non-homogeneous term in Eq.~(\ref{eq:non-homog-term-Rabi}) and plugging it in to Eq.~(\ref{eq:A^(1)-ansatz}), we can write the integral more compactly as
\begin{equation}
  \label{eq:A^(1)-more-compact}
  A^{(1)} =
  \begin{aligned}[t]
    &\i q_{12}\int_0^t\d s \e^{-\i\omega s} \left[\langle\zeta\rangle + 2H_{12}(s)\langle\xi^*\rangle + 2H_{24}(s)v\tilde{I}^{(0,0)}\right] \\
  -&\i q_{21}\int_0^t\d s \e^{\i\omega s}\left[\langle\xi^*\rangle + 2H_{12}(s)\langle\zeta\rangle + 2H_{13}(s)w\tilde{I}^{(0,0)}\right]\,.
  \end{aligned}
\end{equation}
Here we have defined the integral $\tilde{I}^{(a,b)}$, which is used to compute the moments $\langle\zeta\rangle$ and $\langle\xi^*\rangle$. $\tilde{I}^{(a,b)}$ is nothing but the Gaussian integral arising from Eq.~(\ref{eq:A^(1)-ansatz}) with an added ``source" term $\e^{a\xi^* + b\zeta}$:
\begin{multline}
  \label{eq:I-tilde(a,b)-definition}
  \tilde{I}^{(a,b)} \equiv \frac{1}{\pi^2}\int\d^2\xi\d^2\zeta \e^{-|\xi|^2 - |\zeta|^2}\e^{a\xi^* + b\zeta}\\
  \times G(t-s;z,z^*|\xi,\zeta^*)G(s;\xi^*,\zeta|w,v)\,.
\end{multline}
The source term allows us to find the moments as:
\begin{subequations}
  \begin{equation}
    \label{eq:mom-xi*}
    \langle\xi^*\rangle = \frac{\partial}{\partial a}\tilde{I}^{(a,b)}\Bigg|_{a=b=0}
  \end{equation}
  \begin{equation}
    \label{eq:mom-zeta}
    \langle\zeta\rangle = \frac{\partial}{\partial b}\tilde{I}^{(a,b)}\Bigg|_{a=b=0}
  \end{equation}
\end{subequations}
Therefore, getting a solution to \eqref{eq:A^(1)-more-compact} amounts to evaluating the integral $\tilde{I}^{(a,b)}$. Indeed, by exploiting the fact that $\tilde{I}^{(0,0)} = G(t;z,z^*|w,v)$, we can straightforwardly obtain the solution of the integral over $s$ in Eq.~(\ref{eq:A^(1)-more-compact}). Following this procedure we obtain the expression in Eq.~(\ref{eq:A^(1)-sol}), where the time-dependent coefficients $X_i^{(A)}$ are given by
\begin{subequations}
  \begin{align}
    X_1^{(A)}(t) &= \frac{2\e^{-\i\omega t}\left[\bar{n}f_+(t) + \psi_+\left(1-\e^{-\frac{1}{2}\psi_+^*t}\right) \right]}{\left[1 + \bar{n}\left(1-\e^{-\gamma t}\right)\right]|\psi_+|^2}\,, \label{eq:CA_1} \\
    X_2^{(A)}(t) &= \frac{2\bar{n}\e^{-\i\omega t}f_-(t)}{\left[1 + \bar{n}\left(1-\e^{-\gamma t}\right)\right]|\psi_-|^2}\,, \label{eq:CA_2} \\
    X_3^{(A)}(t) &= \frac{2(1+\bar{n})f_+^*(t)}{\left[1 + \bar{n}\left(1-\e^{-\gamma t}\right)\right]|\psi_+|^2}\,, \label{eq:CA_3} \\
    X_4^{(A)}(t) &= \frac{2\left[\bar{n}f_-^*(t) + \psi_-^*\left(1 - \e^{-\frac{1}{2}\psi_-t}\right)\right]}{\left[1 + \bar{n}\left(1-\e^{-\gamma t}\right)\right]|\psi_-|^2} \label{eq:CA_4}\,,
  \end{align}
\end{subequations}
with $f_\pm(t) = \e^{-\gamma t}\psi_\pm^* + \psi_\pm - 2\gamma\e^{-\frac{1}{2}\psi_\pm^*t}$ and $\psi_\pm = \gamma + 2\i(\omega\pm\Omega)$.

Next we consider the case for $B^{(1)}(t;z,z^*|w,v)$, which satisfies a differential equation that is very similar to Eq.~(\ref{eq:linear-correction-Rabi})
\begin{multline}
  \label{eq:Rabi-first-ord-DE-B}
  \dot{B}^{(1)}(t;z,z^*|w,v) = \left[\mathcal{D}(z,z^*) - \i\omega\right]B^{(1)}(t;z,z^*|w,v) \\
  + \mathcal{N}^{(B)}(t;z,z^*|w,v)\,,
\end{multline}
with the non-homogeneous term now given by
\begin{multline}
  \mathcal{N}^{(B)}(t;z,z^*|w,v) = \i\bigg[q_{11}\left(z^* + \frac{\partial}{\partial z^*}\right) - q_{22}\left(z + \frac{\partial}{\partial z}\right)\bigg]\\
  \times G(t;z,z^*|w,v)\,.
\end{multline}
The extra $-\i\omega$ -term in the differential equation (\ref{eq:Rabi-first-ord-DE-B}) requires modifying the ansatz (\ref{eq:A^(1)-ansatz}) by adding an extra factor $\e^{-\i\omega(t-s)}$ into the integrand. Then, the procedure follows exactly the same lines as for  $A^{(1)}(t;z,z^*|w,v)$. We obtain the solution in Eq.~(\ref{eq:B^(1)-sol}), where the time dependent coefficients $C_i^{(B)}$ are given by
\begin{widetext}
  \begin{subequations}
    \begin{align}
      X_1^{(B)}(t) &=  \frac{2\left[\left[\bar{n}\left(q_{11}-q_{22}\right) + q_{11}\right]\psi_-^* + \bar{n}\e^{-\gamma t}\left(q_{11} - q_{22}\right)\psi_-- \e^{-\frac{1}{2}\psi_-t}\left[q_{11}\psi_-^* + 2\gamma\bar{n}\left(q_{11} - q_{22}\right)\right]\right]}{\left[1+\bar{n}\left(1-\e^{-\gamma t}\right)\right]|\psi_-|^2}\,, \label{eq:CB_1} \\
      X_2^{(B)}(t) &=  \frac{2\left[\left[\bar{n}\left(q_{11}-q_{22}\right) - q_{22}\right]\psi_+^* + \bar{n}\e^{-\gamma t}\left(q_{11} - q_{22}\right)\psi_++ \e^{-\frac{1}{2}\psi_+t}\left[q_{22}\psi_+^*-2\gamma\bar{n}\left(q_{11}-q_{22}\right)\right]\right]}{\left[1+\bar{n}\left(1-\e^{-\gamma t}\right)\right]|\psi_+|^2}\,, \label{eq:CB_2} \\
      X_3^{(B)}(t) &=  \frac{2\e^{-\i\omega t}\left[\left[\bar{n}\left(q_{11}-q_{22}\right) + q_{11}\right]\e^{-\gamma t}\psi_-^* + \left(1+\bar{n}\right)\left(q_{11} - q_{22}\right)\psi_-+ \e^{-\frac{1}{2}\psi_-^*t}\left[q_{22}\psi_--2\gamma\bar{n}\left(q_{11}-q_{22}\right)-2\gamma q_{11}\right]\right]}{\left[1+\bar{n}\left(1-\e^{-\gamma t}\right)\right]|\psi_-|^2}\,, \label{eq:CB_3}
    \end{align}
    \begin{align}
      X_4^{(B)}(t) &= \frac{2\e^{-\i\omega t}\left[\left[\bar{n}\left(q_{11}-q_{22}\right) - q_{22}\right]\e^{-\gamma t}\psi_+^* + \left(1+\bar{n}\right)\left(q_{11} - q_{22}\right)\psi_+- \e^{-\frac{1}{2}\psi_+^*t}\left[q_{11}\psi_+ + 2\gamma\bar{n}\left(q_{11}-q_{22}\right) - 2\gamma q_{22}\right]\right]}{\left[1+\bar{n}\left(1-\e^{-\gamma t}\right)\right]|\psi_+|^2} \label{eq:CB_4}\,.
    \end{align}
  \end{subequations}
\end{widetext}
Obtaining the coefficients for the linear corrections $C^{(1)}(t;z,z^*|w,v)$ and $D^{(1)}(t;z,z^*|w,v)$ is straightforward: For $C^{(1)}(t;z,z^*|w,v)$ the coefficients are the same as in Eqs.~(\ref{eq:CB_1}) to (\ref{eq:CB_4}) but with the changes $q_{11} \to q_{22}$, $q_{22} \to q_{11}$, $\omega \to -\omega$. To get coefficients for $D^{(1)}(t;z,z^*|w,v)$ we change from Eqs.~(\ref{eq:CA_1}) to (\ref{eq:CA_4}) $\omega \to -\omega$.

\subsection{Details on computing the dynamics with thermal initial state}

Having now computed the linear corrections to the propagators, we can apply them to compute, as an example, the time evolution of a system where the boson starts out in a thermal state. We can write the Bargmann space representation of the density matrix as a power series in $g$ up to first order
\begin{equation}
  F_\text{B}\left(\rho(t); z, z^*\right) = F_\text{B}^{(0)}\left(\rho(t); z, z^*\right) + gF_\text{B}^{(1)}\left(\rho(t); z, z^*\right)\,,
\end{equation}
where the terms on the right hand side are given as the integration of the propagators acting on the initial state, as shown in Eq.~(\ref{eq:rho-in-holomorphic_main}).

For the unperturbed term $F_\text{B}^{(0)}\left(\rho(t); z, z^*\right)$ the integration yields the holomorphic representation of the thermal state multiplied by the unperturbed dynamics of the qubit. For the first order correction we need to again compute a complex Gaussian integral with the initial state as shown in Eq.~(\ref{eq:rho-in-holomorphic_main}). The integrand becomes of the same form as in the brackets of Eq.~(\ref{eq:A^(1)-more-compact}), so the procedure is equivalent to the one for computing the expression for $A^{(1)}(t;z,z^*|w,v)$. We define a new integral with additional source term $\e^{k_1w+k_2v}$ and then obtain the moments $\langle w\rangle$ and  $\langle v\rangle$ via differentiation. These moments will give contributions that are linear in $z$ and $z^*$, such that in the end we obtain Eq.~(\ref{eq:Rabi-solution-linear-in-g2}) where the time dependent coefficients are given by
\begin{subequations}
  \begin{align}
    \begin{split}
      \text{coeff}^{(A)}(t) &= \frac{2q_{12}\bar{n}\e^{-\i\omega t}\left(1 - \e^{-\frac{1}{2}\psi_-^*t}\right)}{(1+\bar{n})\psi_-^*} \\
                            &\quad- \frac{2q_{21}\e^{\i\omega t}\left(1 - \e^{-\frac{1}{2}\psi_+t}\right)}{\psi_+}\,, \label{eq:F_B11-corr_coeffA}
    \end{split}\\
    \text{coeff}^{(B)}(t) &= \frac{2\left(1 - \e^{-\frac{1}{2}\psi_+t}\right)\left(\bar{n}\left(q_{11} - q_{22}\right) - q_{22}\right)}{(1+\bar{n})\psi_+} \,, \label{eq:F_B12-corr_coeffBz} \\
    \text{coeff}^{(C)}(t) &= -\frac{2\left(1 - \e^{-\frac{1}{2}\psi_-^*t}\right)\left(\bar{n}\left(q_{11} - q_{22}\right) + q_{11}\right)}{(1+\bar{n})\psi_-^*} \,, \label{eq:F_B21-corr_coeffCz}
  \end{align}
  \begin{align}
    \begin{split}
      \text{coeff}^{(D)}(t) &= \frac{2q_{21}\bar{n}\e^{\i\omega t}\left(1 - \e^{-\frac{1}{2}\psi_+t}\right)}{(1+\bar{n})\psi_+} \\
                            &\quad- \frac{q_{12}\e^{-\i\omega t}\left(1 - \e^{-\frac{1}{2}\psi_-^*t}\right)}{\psi_-^*}\,. \label{eq:F_B22-corr_coeffD}
    \end{split}
  \end{align}
\end{subequations}
Here, again, $\psi_\pm = \gamma + 2\i(\omega\pm\Omega)$.

\section{Quantum Rabi model: second order calculation details} \label{sec:quantum-rabi-model-sec-ord}

For the second order perturbation theory calculations in section \ref{sec:pert-solut-quant}, we assume for simplicity that the initial state of the boson is thermal. Therefore, we look for a solution to elements of the density matrix in Bargmann space  $\left[F_\text{B}\left(\rho(t); z,z^*\right)\right]_{ij} \equiv f_{ij}$. The equations for the matrix elements are given by
\begin{subequations}
  \begin{align}
    \dot{f}_{11}^{(2)} &=
                         \begin{aligned}[t]
                           &\mathcal{D}(z,z^*)f_{11}^{(2)} \\
                           &+ \i \left[\left(z^* + \frac{\partial}{\partial z^*}\right)f_{12}^{(1)} - \left(z + \frac{\partial}{\partial z}\right)f_{21}^{(1)}\right]\,, \label{eq:Rabi-2ord-11}
                         \end{aligned} \\
    \dot{f}_{12}^{(2)} &=
                         \begin{aligned}[t]
                           \big[&\mathcal{D}(z,z^*) - \i\omega\big]f_{12}^{(2)} \\
                                &+ \i \left[\left(z^* + \frac{\partial}{\partial z^*}\right)f_{11}^{(1)} - \left(z + \frac{\partial}{\partial z}\right)f_{22}^{(1)}\right]\,,
                         \end{aligned} \\
    \dot{f}_{21}^{(2)} &=
                         \begin{aligned}[t]
                           \big[&\mathcal{D}(z,z^*) + \i\omega\big]f_{21}^{(2)} \\
                                &+ \i \left[\left(z^* + \frac{\partial}{\partial z^*}\right)f_{22}^{(1)} - \left(z + \frac{\partial}{\partial z}\right)f_{11}^{(1)}\right]\,, 
                         \end{aligned} \\
    \dot{f}_{22}^{(2)} &=
                         \begin{aligned}[t]
                           &\mathcal{D}(z,z^*)f_{22}^{(2)} \\
                           &+ \i \left[\left(z^* + \frac{\partial}{\partial z^*}\right)f_{21}^{(1)} - \left(z + \frac{\partial}{\partial z}\right)f_{12}^{(1)}\right]\,. \label{eq:Rabi-2ord-22}
                         \end{aligned}
  \end{align}
\end{subequations}
Here the nonhomogeneous terms contain solutions of the first order corrections $f_{ij}^{(1)}$, which are known and can be obtained from Eq.~(\ref{eq:Rabi-solution-linear-in-g2}).

Based on the knowledge of the zeroth and first order solutions to the Rabi problem, we can make an ansatz for the second order solution and say that it should be a polynomial that is at most of the second order in the complex variables $z$ and $z^*$. The ansatz then reads:
\begin{multline}
  \label{eq:Rabi-2nd-ord-ansatz}
  f_{ij}^{(2)}(t;z,z^*) = \Big[x^{(2,0)}_{ij}(t)z^2 + x^{(1,1)}_{ij}(t)zz^* + x^{(0,2)}_{ij}(t)z^{*2} \\
    + x^{(1,0)}_{ij}(t)z + x^{(0,1)}_{ij}(t)z^* + x^{(0,0)}_{ij}(t)\Big]\e^{\frac{\bar{n}}{1+\bar{n}}zz^*}\,.
\end{multline}

Here $x_{ij}(t)$ are some time-dependent functions that can be obtained by plugging the ansatz into the second order DEs (\ref{eq:Rabi-2ord-11}) to (\ref{eq:Rabi-2ord-22}). Note that the dependency of the ansatz on $z, z^*$ is polynomial, when the thermal state initial condition of the boson is factored out.

Using the ansatz in Eqs.~(\ref{eq:Rabi-2ord-11}) to (\ref{eq:Rabi-2ord-22}) and solving the resulting differential equations for $x_{ij}^{(n,m)}$, we notice that $x^{(1,0)}_{ij} = x^{(0,1)}_{ij} = 0$. Hence we obtain Eq.~(\ref{eq:Rabi-2nd-ord-sol}). However, we must note that this solution has an issue: although the solution works for small times, the coefficients $x_{ij}^{(0,0)}$ contain a term that depends linearly on time (see Eqs.~(\ref{eq:x_11^00}) and (\ref{eq:x_12^00})). These \textit{secular} terms arise due to application of perturbation theory and they diverge for late times, leading to the breakdown of the solution \cite{Verhulst2006}. To cure this issue, we apply multiscale perturbation theory at the second order.

\subsection{Multiscale perturbation theoretic treatment}
In multiscale perturbation theory we assume that there exist two separate timescales; a ``fast'' one $t$ and a ``slow'' one $\tau = g^2t$, which becomes significant only at large times $t\propto g^{-2}$. The time scales are assumed to be separate variables \cite{Strogatz2018}. Therefore we write the solution as $f(t, \tau)$, which gives the following contributions when differentiated with respect to time $t$:
\begin{equation}
  \frac{\d}{\d t}f(t, \tau) = \frac{\partial f}{\partial t} + g^2\frac{\partial f}{\partial \tau}\,.
\end{equation}
Applying the regular perturbation series expansion for $f(t, \tau)$ and collecting the terms of the same order of $g$, we notice that the zeroth and first order equations are not modified with respect to standard perturbation theory. In contrast, the second order equation gains the extra derivative term with respect to the slow time $\tau$. This means that the left-hand side of Eqs.~(\ref{eq:Rabi-2ord-11}) to (\ref{eq:Rabi-2ord-22}) transforms to
\begin{equation}
  \label{eq:time-deriv-trafo}
  \dot{f}_{ij}^{(2)} = \frac{\partial}{\partial \tau}f_{ij}^{(0)}(t, \tau;z,z^*) + \frac{\partial}{\partial t}f_{ij}^{(2)}(t, \tau; z,z^*)\,.
\end{equation}
Note that the extra differentiation with respect to the slow time $\tau$ acts on the zeroth order solution $f_{ij}^{(0)}$. Next we insert the known solution for the unperturbed case (the first term of Eq.~(\ref{eq:Rabi-solution-linear-in-g2})) and promote the initial data of the qubit $q_{ij}$ to become functions of the slow time $\tau$, i.e. $q_{ij} \to Q_{ij}(\tau)$ \cite{Strogatz2018}. 

In order to remove the problematic secular terms in the solutions to the second order differential equations (\ref{eq:Rabi-2ord-11}) to (\ref{eq:Rabi-2ord-22}), we apply the above promotion to every qubit initial data in the differential equations. Then we isolate the expressions in the resulting PDEs that would cause the appearance of the secular term, and treat them  as a new independent differential equation for the promoted data $Q_{ij}(\tau)$. Setting this PDE to zero and solving for $Q_{ij}(\tau)$ removes the secular term.

\ant{Having solved for the functions $Q_{ij}(\tau)$ (see Eqs.~(\ref{eq:Q11(T)}) and (\ref{eq:Q12(T)})), we proceed with solving for the time dependent functions $\tilde{x}_{ij}^{(n,m)}(t, \tau)$ from Eq.~(\ref{eq:2nd-ord-ms-expsigmaz}), where the multiscaling has been taken into account.}

\begin{widetext}
  \subsection{Form of the coefficients $x_{ij}^{(a,b)}$ and $Q_{ij}(T)$}
  For all expressions below we have again that $\psi_\pm = \gamma + 2\i(\omega\pm\Omega)$. Recall that these solutions are valid when $g\lessapprox\Delta$.

  \begin{subequations}
    \begin{align}
      \begin{split}
        x_{11}^{(0,0)}(t) &= \left[\frac{8\e^{-\frac{1}{2}\psi_+t} \gamma \left[(\bar{n}+1)q_{22}-\bar{n} q_{11}\right]}{\psi_+|\psi_+|^2} - \frac{4 \e^{-\frac{1}{2}\psi_-t} (2 \bar{n} \gamma +\psi_-^*) \left[(\bar{n}+1) q_{11}-\bar{n} q_{22}\right]}{(\bar{n}+1) \psi_- |\psi_-|^2} + \text{h.c.}\right] \\
                          &\qquad+ \frac{\e^{-\gamma t} \left[8 \bar{n} (\bar{n}+1) \left(\gamma^2+4 \left(\omega ^2+\Omega ^2\right)\right) q_{11}-4 \left[(2\bar{n}^2 + 2\bar{n} + 1)|\psi_-|^2 + 16\bar{n}^2\omega\Omega\right]q_{22}\right]}{(\bar{n}+1)|\psi_+|^2|\psi_-|^2} \\
                          &\qquad+ \frac{4\gamma \left[q_{22} \left((2\bar{n}+1)|\psi_-|^2 + 16\bar{n}\omega\Omega\right)- q_{11}\left[(2\bar{n}+1)|\psi_+|^2 - 16\bar{n}\omega\Omega\right]\right]}{|\psi_+|^2|\psi_-|^2}t \\
                          &\qquad+ 4q_{11} \left[\frac{4\gamma^2\bar{n}}{|\psi_+|^4}+\frac{4\gamma^2(\bar{n}+1)}{|\psi_-|^4} - \frac{\bar{n}+2}{|\psi_-|^2} - \frac{\bar{n}}{|\psi_+|^2}\right] - 4q_{22}\left[\frac{4\gamma^2\bar{n}}{|\psi_-|^4} + \frac{4\gamma^2(\bar{n}+1)}{|\psi_+|^4} - \frac{\bar{n} (\bar{n}+2)}{(\bar{n}+1)|\psi_-|^2} - \frac{\bar{n}+1}{|\psi_+|^2}\right]\,,
      \end{split}
      \label{eq:x_11^00}
      \\[2ex]
      \begin{split}
        x_{11}^{(1,1)}(t) &= \left[\frac{4\left[\bar{n}q_{11} - (\bar{n}+1)q_{22}\right]\e^{-\frac{1}{2}\psi_+t}}{(\bar{n}+1)|\psi_+|^2} + \frac{4\bar{n}\left[(\bar{n}+1) q_{11}-\bar{n} q_{22}\right]\e^{-\frac{1}{2}\psi_-t}}{(\bar{n}+1)^2 |\psi_-|^2} + \text{h.c.}\right] \\
                          &\qquad- \frac{\left(1 + \e^{-\gamma t}\right)\left[8\bar{n} (\bar{n}+1)\left(\gamma^2+4 \left(\omega^2+\Omega^2\right)\right)q_{11} - 4\left[(2\bar{n}^2 + 2\bar{n} + 1)|\psi_-|^2 + 16\bar{n}^2\omega\Omega\right]q_{22}\right]}{(\bar{n}+1)^2|\psi_+|^2|\psi_-|^2}\,, 
      \end{split}
      \\[2ex]
      \begin{split}
        x_{11}^{(2,0)}(t) &= \frac{\left(4 (\bar{n}+1) q_{11}-4 \bar{n}q_{22}\right) \e^{-\frac{1}{2}\psi_-^*t}}{(\bar{n}+1)\psi_+\psi_-^*} + \frac{4 \bar{n} \left(\bar{n}q_{11}-(\bar{n}+1) q_{22}\right) \e^{-\frac{1}{2}\psi_+t}}{(\bar{n}+1)^2\psi_+\psi_-^*} \\
                          &\qquad+ \frac{\e^{-(\gamma + 2\i\Omega)t} \left[4 \bar{n}(\bar{n}+1)(\gamma + 2\i\Omega)q_{22} - 2\left[(2\bar{n}^2 + 2\bar{n} + 1)\psi_-^* + 4\i \bar{n}^2\omega\right]q_{11}\right]}{(\bar{n}+1)^2 (\gamma + 2\i\Omega)\psi_+\psi_-^*} \\
                          &\qquad+ \frac{4 \bar{n} (\bar{n}+1)(\gamma + 2\i\Omega)q_{22} - 2\left[(2\bar{n}^2 + 2\bar{n} + 1)\psi_+ - 4\i \bar{n}^2\omega\right]q_{11}} {(\bar{n}+1)^2 (\gamma + 2\i\Omega) \psi_+\psi_-^*}\,,
      \end{split}
      \\[2ex]
      x_{11}^{(0,2)}(t) &= x_{11}^{(2,0)*}(t)\,,
      \\[2ex]
      \begin{split}
        x_{12}^{(0,0)}(t) &= -\frac{2\i \e^{\i t \omega } \left(\gamma +2 \gamma  \bar{n}^2+3 \gamma  \bar{n}+2\i \bar{n} \omega \right)q_{21}}{(\bar{n}+1) \omega \psi_+\psi_-} + \frac{\e^{-(\gamma + \i\omega)t} \left[-4\gamma\left(2\bar{n}^2 + 2\bar{n} + 1\right)q_{21} + 8\bar{n}(\bar{n}+1)(\gamma + 2\i\omega)q_{12}\right]}{\gamma  (\bar{n}+1)\psi_+\psi_-} \\
                          &\qquad+ \frac{\e^{-\frac{1}{2}(\gamma + 2\i\Omega)t} \left[4 \psi_-^*\left[2\gamma(2\bar{n}^2 + 2\bar{n} + 1) + \bar{n}\psi_-\right]q_{21} - 4\psi_+(\bar{n}+1)(4\gamma \bar{n} + \psi_-)q_{12}\right]}{(\bar{n}+1)|\psi_-|^2\psi_+\psi_-^*} \\
                          &\qquad+ \frac{\e^{-\frac{1}{2}(\gamma - 2\i\Omega)t} \left[4\psi_+^*\left[2\gamma(2\bar{n}^2 + 2\bar{n} + 1) + \bar{n}\psi_+\right]q_{21} - 4\psi_-(\bar{n}+1)(4\gamma \bar{n} + \psi_+)q_{12}\right]}{(\bar{n}+1) |\psi_+|^2\psi_+^*\psi_-} \\
                          &\qquad+ \e^{-i\omega t}\Bigg[\frac{2\psi_+^*\psi_-^*\gamma(2\bar{n}+1)(2\omega + \i\gamma)q_{21} + 8 \omega\left[(\gamma - 2\i\omega)^2(\gamma + 3\gamma \bar{n} - 2\i \bar{n}\omega) - 4\Omega^2(\gamma + \gamma \bar{n} + 2\i \bar{n}\omega)\right]q_{12}}{\gamma  \omega  \left(\psi_+^*\psi_-^*\right)^2} \\
                          &\qquad- \frac{4(2\bar{n}+1)(\gamma - 2\i\omega)q_{12}}{\psi_+^*\psi_-^*}t\Bigg]\,,
      \end{split}
      \label{eq:x_12^00}
      \\[2ex]
      \begin{split}
        x_{12}^{(1,1)}(t) &= \frac{4\e^{\i\omega t}\left(2\bar{n}^2 + 2\bar{n} + 1\right)q_{21}}{(\bar{n}+1)^2\psi_+^*\psi_-^*} - \frac{8\e^{-\i\omega t}\bar{n}(\gamma - 2\i\omega )q_{12}}{\gamma(\bar{n}+1)\psi_+^*\psi_-^*} + \frac{\e^{-(\gamma + \i\omega)t} \left[4 \gamma  \left(2 \bar{n}^2+2\bar{n}+1\right) q_{21}-8 \bar{n} (\bar{n}+1)(\gamma +2\i\omega)q_{12}\right]}{\gamma
                            (\bar{n}+1)^2 \psi_+^*\psi_-^*} \\
                          &\qquad+ \frac{\e^{-\frac{1}{2}(\gamma +2\i\Omega )t} \left[8\psi_+\bar{n}(\bar{n}+1)q_{12} - 4\psi_-^*\left(2 \bar{n}^2+2 \bar{n}+1\right)q_{21}\right]}{(\bar{n}+1)^2 \psi_+^*\psi_-^{*2}} \\
                          &\qquad+ \frac{\e^{-\frac{1}{2}(\gamma -2\i\Omega)t} \left[8\psi_-\bar{n}(\bar{n}+1)q_{12} - 4\psi_+^*\left(2 \bar{n}^2+2\bar{n}+1\right)q_{21}\right]}{(\bar{n}+1)^2\psi_+^{*2}\psi_-^*}\,,
      \end{split}
    \end{align}

    \begin{align}
      \begin{split}
        x_{12}^{(2,0)}(t) &= -\frac{2\e^{-\i\omega t}(2\bar{n}^2 + 2\bar{n} + 1)q_{12}}{(\bar{n}+1)^2 (\gamma +2\i\Omega) \psi_-^*} + \frac{4\e^{\i\omega t} \bar{n} q_{21}}{(\bar{n}+1)\psi_+^2} + \frac{4 \e^{-\frac{1}{2}(\gamma +2\i\Omega)t} \left[-2\psi_-^*\bar{n} (\bar{n}+1)q_{21} + \psi_+(2\bar{n}^2 + 2\bar{n} + 1)q_{12}\right]}{(\bar{n}+1)^2 \psi_-^*\psi_+^2} \\
                          &\qquad+ \frac{2 \e^{-(\psi_+ + \i\Omega)t} \left[2 \bar{n} (\bar{n}+1) (\gamma +2\i\Omega)q_{21} - \psi_+(2\bar{n}^2 + 2\bar{n} + 1)q_{12} \right]}{(\bar{n}+1)^2 (\gamma +2\i\Omega)\psi_+^2}\,,
      \end{split}
      \\[2ex]
      x_{12}^{(0,2)}(t) &= -\frac{2\e^{-\i \omega t}(2\bar{n}^2 + 2\bar{n} + 1)q_{12}}{(\bar{n}+1)^2 (\gamma -2\i\Omega)\psi_+^*} + \frac{4\e^{\i \omega t} \bar{n} q_{21}}{(\bar{n}+1) \psi_-^2} + \frac{4 \e^{-\frac{1}{2}(\gamma -2 \i \Omega)t} \left[-2\psi_+^*\bar{n}(\bar{n}+1) q_{21} + \psi_-(2 \bar{n}^2 + 2\bar{n} + 1)q_{12}\right]}{(\bar{n}+1)^2 \psi_+^*\psi_-^2} \\
                          &\qquad+ \frac{2 \e^{-(\psi_- - \i\Omega)t} \left[2 \bar{n} (\bar{n}+1)(\gamma -2\i\Omega) q_{21} - \psi_-(2\bar{n}^2 + 2\bar{n} + 1) q_{12}\right]}{(\bar{n}+1)^2 (\gamma -2\i\Omega)\psi_-^2}\,. \label{eq:x_12^02}
    \end{align}
  \end{subequations}
  For the other elements we have that $x_{21}^{(0,0)}(t) = x_{12}^{(0,0)*}(t)$, $x_{21}^{(1,1)}(t) = x_{12}^{(1,1)*}(t)$, $x_{21}^{(2,0)}(t) = x_{12}^{(0,2)*}(t)$ and $x_{21}^{(0,2)}(t) = x_{12}^{(2,0)*}(t)$. The coefficients $x_{22}^{(a,b)}(t)$ can be obtained from $x_{11}^{(a,b)}(t)$ by applying the changes $q_{11} \to q_{22}$, $q_{22} \to q_{11}$ and $\Omega \to -\Omega$.

  Next we write the corrected zeroth order expressions for the qubit density matrix elements that were obtained by the application of multiscale perturbation theory.

  \begin{subequations}
    \begin{align}
      \begin{split}
      Q_{11}(g^2t) &= \frac{\left(|\psi_+|^2(2\bar{n}+1) - 16\bar{n}\omega\Omega \right)q_{11} - \left(|\psi_-|^2(2\bar{n}+1) + 16\bar{n}\omega\Omega\right)q_{22}}{2 (2 \bar{n}+1) \left(\gamma ^2+4 \left(\omega ^2+\Omega^2\right)\right)}\e^{-\frac{8g^2\gamma  (2 \bar{n}+1) \left(\gamma^2+4 \left(\omega ^2+\Omega ^2\right)\right)}{|\psi_+|^2|\psi_-|^2}t} \\
                   &\qquad+ \frac{|\psi_-|^2(2\bar{n}+1) + 16\bar{n}\omega\Omega}{2 (2 \bar{n}+1) \left(\gamma ^2+4 \left(\omega ^2+\Omega^2\right)\right)}
      \end{split}
      \label{eq:Q11(T)}\\[2ex]
      Q_{12}(g^2t) &= q_{12} \e^{-\frac{4g^2(2 \bar{n}+1) (\gamma -2 i \omega )}{4\Omega ^2+(\gamma -2 i \omega )^2}t} \label{eq:Q12(T)}
    \end{align}
  \end{subequations}
And $Q_{22} = 1 - Q_{11}$ and $Q_{21} = Q_{12}^*$.

\end{widetext}

\bibliography{library}

\end{document}